\documentclass{article}

\usepackage{microtype}
\usepackage{graphicx}
\usepackage{xspace}
\usepackage{booktabs} 
\usepackage{mathtools}
\usepackage{subcaption}
\usepackage{amsmath,amssymb,amsthm}
\usepackage{multirow}
\usepackage{appendix}

\usepackage{hyperref}

\newtheorem{definition}{Definition}
\newtheorem{theorem}{Theorem}
\newtheorem{lemma}{Lemma}
\newtheorem{assume}{Assumption}
\newtheorem{corollary}{Corollary}
\newcommand{\kstar}{^{\textstyle *}}
\DeclareMathOperator*{\argmin}{arg\,min}

\usepackage{icml2020}

\newcommand{\algo}{\texttt{MergeComp}\xspace}

\begin{document}

\twocolumn[
\icmltitle{\algo: A Compression Scheduler for Scalable Communication-Efficient Distributed Training}


\icmlsetsymbol{equal}{*}

\begin{icmlauthorlist}
\icmlauthor{Zhuang Wang}{rice}
\icmlauthor{Xinyu Wu}{rice}
\icmlauthor{T.S. Eugene Ng}{rice}
\end{icmlauthorlist}

\icmlaffiliation{rice}{Department of Computer Science, Rice University, Houston, TX, USA}

\icmlcorrespondingauthor{Zhuang Wang}{Zhuang.Wang@rice.edu}
\icmlcorrespondingauthor{T.S. Eugene Ng}{eugeneng@rice.edu}

\icmlkeywords{Machine Learning, ICML}

\vskip 0.3in
]



\printAffiliationsAndNotice{}  

\begin{abstract}

Large-scale distributed training is increasingly becoming communication bound.
Many gradient compression algorithms have been proposed to reduce the communication overhead and improve scalability.
However, it has been observed that in some cases gradient compression may even harm the performance of distributed training.

In this paper, we propose \algo, a compression scheduler to optimize the scalability of communication-efficient distributed training.
It automatically schedules the compression operations to optimize the performance of compression algorithms without the knowledge of model architectures or system parameters.
We have applied \algo to nine popular compression algorithms.
Our evaluations show that \algo can improve the performance of compression algorithms by up to 3.83$\times$ without losing accuracy.
It can even achieve a scaling factor of distributed training up to 99\% over high-speed networks.
\end{abstract}
\section{Introduction}

Distributed training has been widely adopted to accelerate the model training of deep neural networks (DNN). Data parallelism is a standard strategy to scale out the training~\cite{projectadam, PSosdi2014, horovod}.
Multiple workers (e.g., GPUs) are employed and each worker has a replica of the training model.
The training dataset is divided into multiple partitions so that each worker just takes one partition as its training data. 
Distributed training can significantly reduce the training time. 
However, since it is necessary for the workers to synchronize the gradient updates to ensure a consistent model, it is nontrivial for distributed training to achieve near-linear scalability due to the non-negligible communication overhead~\cite{zhang2020network, MLPerf, bytescheduler, pipedream, luo2018parameter}.

To tackle the communication overhead and improve scalability, a plethora of gradient compression algorithms have been proposed to reduce the volume of transferred data for synchronization.
There are two main types of compression techniques: 1) \textit{sparsification}, which selects a subset of gradients for communication~\cite{strom2015scalable, dgc, tsuzuku2018variance, stich2018sparsified}; 
and 2) \textit{quantization}, which quantizes gradients (e.g., FP32) to fewer bits~\cite{inceptionn, QSGD, 1bit,  efsignsgd, wu2018error}. 
Theoretically, gradient compression can dramatically reduce the communication overhead thanks to the much smaller communicated data size~\cite{alistarh2018convergence, jiang2018linear}.

\vskip -0.04in
However, performing gradient compression to reduce the communicated data size is not free.
Some recent works~\cite{xu2020compressed, SwitchML, li2018network, gupta2020fast} noticed that gradient compression harms the scalability of distributed training in some cases and suggested that these compression techniques are only beneficial for training over slow networks~\cite{3lc}.

\vskip -0.04in
In this paper, we first fully analyze the practical performance of compression algorithms.
Unfortunately, we observe that the performance of training with compression algorithms is far from optimal, and in most cases even worse than training without any compression due to the prohibitive overhead of compression operations.

\vskip -0.04in
We then propose \algo, a compression scheduler to optimize the performance of compression algorithms.
It automatically determines the near-optimal schedule for applying compression operations to reduce the compression overhead for various DNN models.
Meanwhile, it can adaptively overlap the communication with the computation to further reduce the communication overhead for different system parameters (e.g., the number of workers and network bandwidth capacities).


We have applied \algo to nine popular compression algorithms and show extensive experiments on image classification and image segmentation tasks.
It demonstrates that \algo can improve the performance of compression algorithms by up to 3.83$\times$ without losing accuracy.
\algo can even achieve a scaling factor up to 99\% over high-speed networks (e.g., NVLink). 

We open-source \algo and hope this work could pave the way for applying gradient compression algorithms to distributed training in production environments.
\section{Background}

We first introduce the main gradient compression algorithms for distributed training and then describe how they are implemented in existing ML frameworks.

\subsection{Gradient compression algorithms}
\label{sec:compress}
Single precision (also known as FP32) is a common floating point format to represent the weights and gradients in deep learning.
Without any compression, the gradients are communicated in FP32 for synchronization, resulting in non-negligible communication overhead.

There are two main compression techniques to reduce the communicated data size: sparsification and quantization.
Sparsification algorithms select only a subset of the original stochastic gradients for synchronization.
Strom~\cite{strom2015scalable} proposed to communicate gradients larger than a predefined threshold, while the threshold is non-trivial to choose in practice.
Therefore, some sparsification algorithms, such as Rand-k~\cite{stich2018sparsified}, Top-k~\cite{aji2017sparse} and DGC~\cite{dgc}, choose a fixed compression ratio to sparsify the gradients; other sparsification algorithms~\cite{adacomp, sparse2018gradient} automatically tune the compression ratio to control the variance of the gradients.

There are two classes of quantization algorithms: limited-bit and codebook-based.
In limited-bit quantization algorithms, each gradient element is mapped to fewer bits, such as FP16, 8 bits~\cite{8bits}, 2 bits (TernGrad~\cite{wen2017terngrad}) and even 1 bit (OneBit~\cite{1bit}, EFSignSGD~\cite{efsignsgd}, SignSGD~\cite{ signsgd}, SigNUM~\cite{signum} and dist-EF-SGD~\cite{dist-ef-sgd}).
In codebook-based quantization algorithms, such as QSGD~\cite{QSGD} and ECQ-SGD~\cite{wu2018error}, gradients are randomly rounded to a discrete set of values that preserve the statistical properties of the original stochastic gradients.

\subsection{Compression in existing ML frameworks}

\begin{figure}[t!]
\begin{center}
\centerline{\includegraphics[width=0.95\columnwidth]{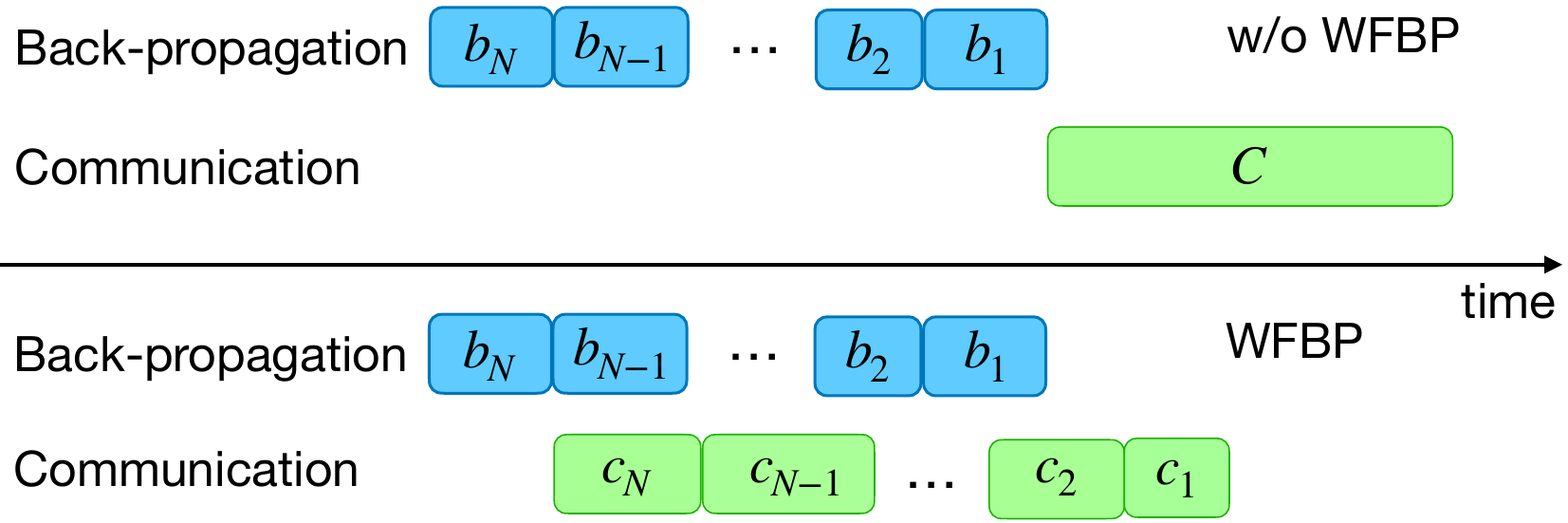}}
\vskip -0.1in
\caption{Distributed training without and with WFBP.}
\label{fig:wfbp}
\end{center}
\vskip -0.4in
\end{figure}

A DNN model consists of a number of layers. 
The distributed training involves three steps: 1) forward propagation, which takes as input a batch of training data, propagates it through the model, and calculates the loss function; 2) back-propagation, which computes the gradients of the parameters with the loss value layer by layer; 3) synchronization, which aggregates the gradient updates from all the workers and update the model.

Since back-propagation is performed layer by layer, the gradients of layer $l$ could be communicated during the computation of the gradients in layer $l-1$, as shown in Figure~\ref{fig:wfbp}.
In other words, the communication begins before the completion of the computation. 
This wait-free back-propagation (WFBP) mechanism significantly reduces the communication overhead by overlapping the communication with the computation~\cite{poseidon, horovod, pytorch_ddp, bytescheduler}.

Because of WFBP, existing distributed ML frameworks, such as PyTorch~\cite{pytorch}, TensorFlow~\cite{tensorflow}, and Horovod~\cite{horovod}, apply compression algorithms layer by layer for distributed training to further reduce the communication time.
In addition, it is explicitly mentioned in many papers that the compression algorithms are applied in a layer-wise fashion~\cite{efsignsgd, 3lc, wen2017terngrad, adacomp, aji2017sparse}.

\section{The Practical Performance of Compression Algorithms}
\label{sec:measurement}

In this section, we first measure the practical performance of popular compression algorithms for distributed training. 
We then analyze the root cause of their poor performance and discuss the challenge to address it.


\subsection{Performance measurement}
To profile the performance of gradient compression, we empirically measure the training speed of popular compression algorithms, including both sparsification and quantization.
Suppose the training speed with $n$ workers is $T_n$.
The scaling factor~\cite{zhang2020network} is defined as
\vskip -0.2in
\begin{equation*}
\small
    \textup{scaling factor} = \frac{T_n}{nT_1}.
\end{equation*}
\vskip -0.1in

\textbf{Setup.} 
All experiments are conducted on a server equipped with 8 GPUs (NVIDIA Tesla V100 with 32 GB memory), two 20-core/40-thread processors (Intel Xeon Gold 6230 2.1GHz), PCIe 3.0 $\times$16, and NVLink.
The server has an Ubuntu 18.04.4 LTS system and the software environment includes PyTorch-1.7.1, Horovod-0.21.1, CUDA-10.1, Open MPI-4.0.2 and NCCL-2.8.3.
The batch size is 64 and the gradient sparsity of sparsification algorithms is 99\%.

The evaluated schemes and their corresponding communication primitives~\footnote{We observe that NCCL2 allgather receives random data and crashes the training at times. This behavior is also observed in~\cite{xu2020compressed, dgc}. So we only evaluate MPI allgather.} are shown in Table~\ref{table:compress_algo}.
Allreduce is used for FP32 and FP16~\cite{horovod, pytorch_ddp, NCCL};
allgather is used for other schemes because allreduce does not support sparse tensors and requires input tensors to be of the same data type and dimension for reduction~\cite{xu2020compressed, sparcml}.


\begin{figure}[t!]
\begin{center}
\centerline{\includegraphics[width=0.9\columnwidth]{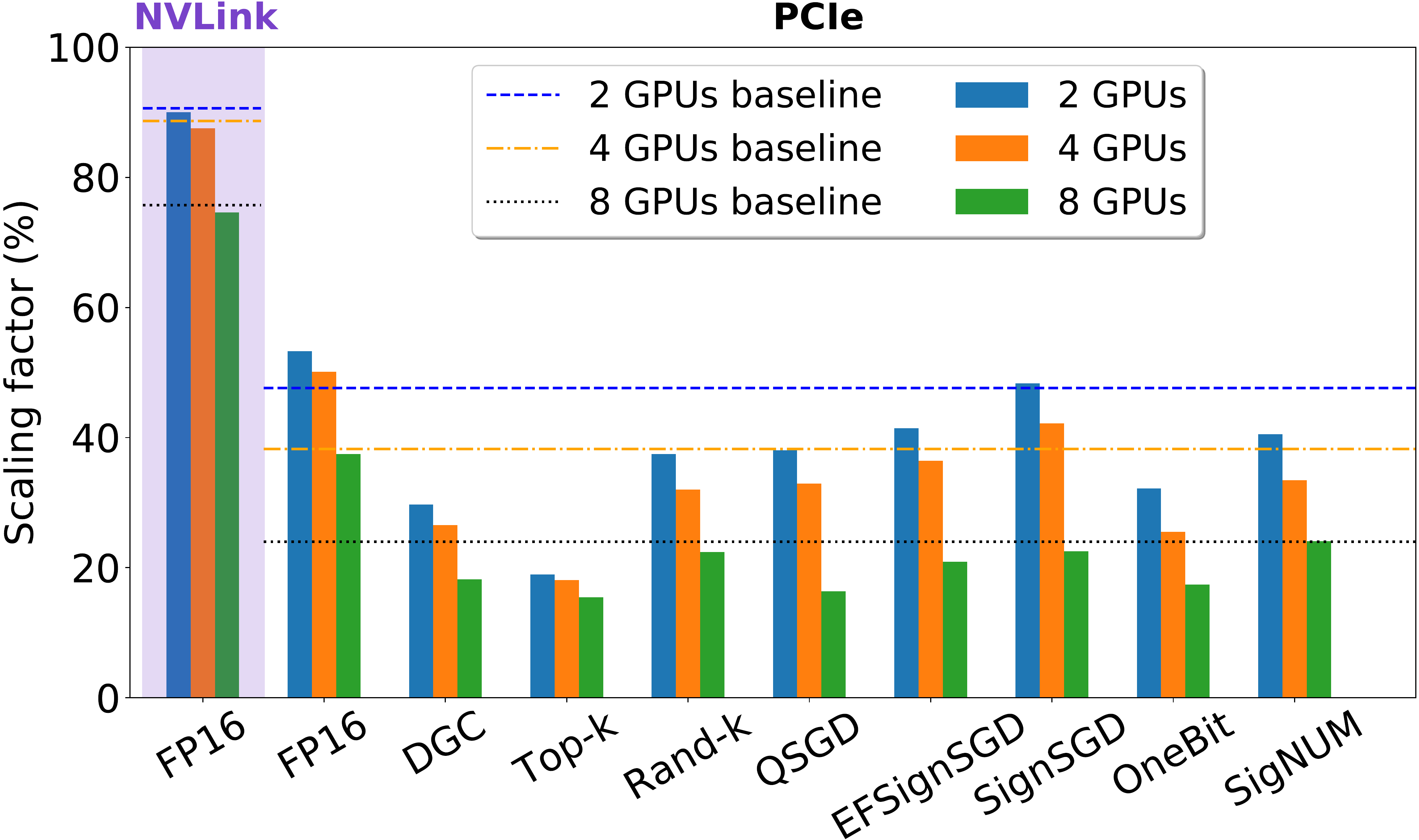}}
\vskip -0.1in
\caption{The scaling factors of ResNet50 on CIFAR10 with various compression algorithms, which apply layer-wise compression. It shows that compression algorithms do not scale well.}
\label{fig:resnet50_lw}
\vskip -0.3in
\end{center}
\end{figure}

\begin{figure*}[t!]
    \centering
    \begin{subfigure}[t]{0.32\linewidth}
	\includegraphics[width=\linewidth]{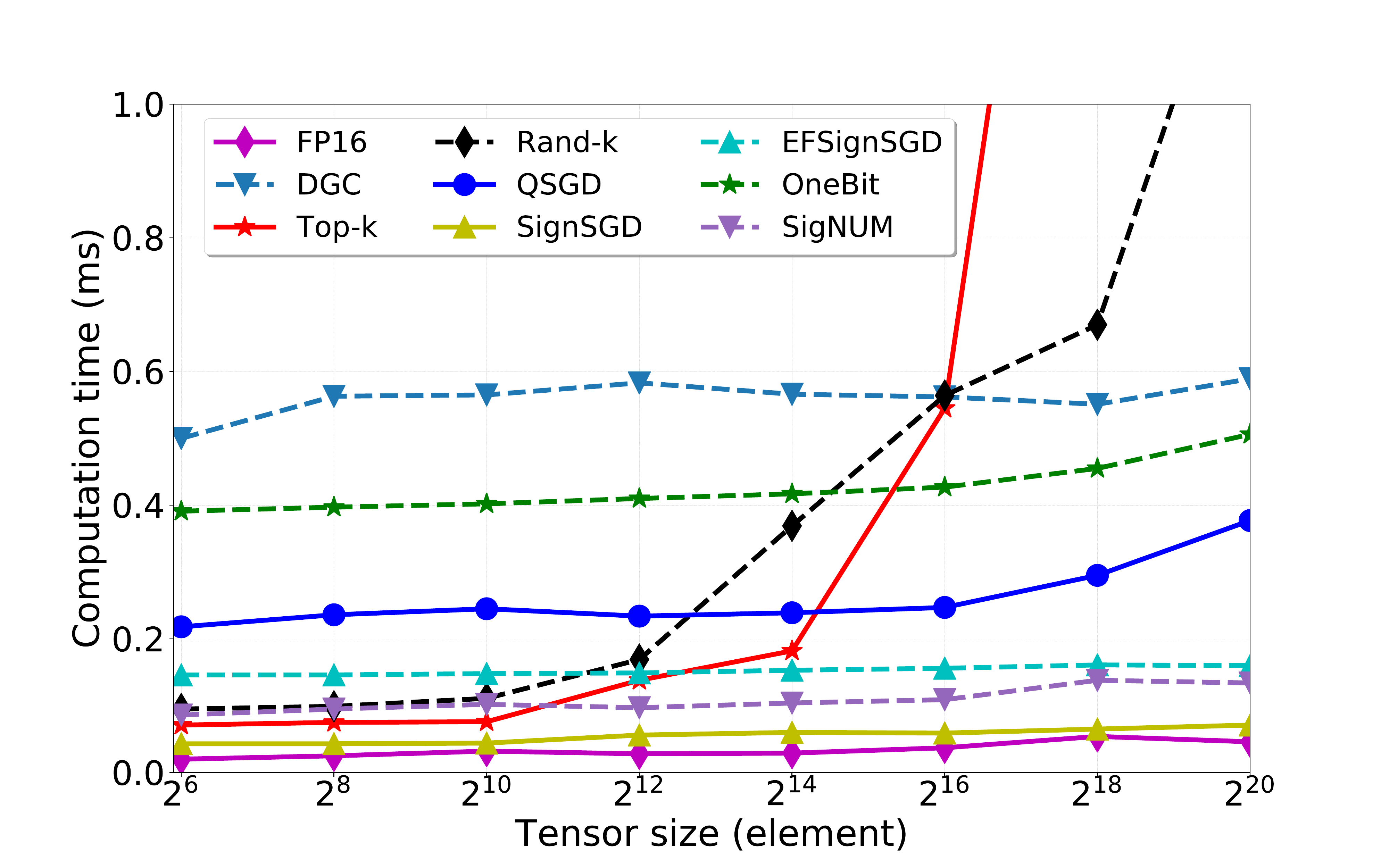}
	\caption{Encoding overhead}
	\label{fig:comp_overhead}
    \end{subfigure}    
	\begin{subfigure}[t]{0.32\linewidth}
        \includegraphics[width=\linewidth]{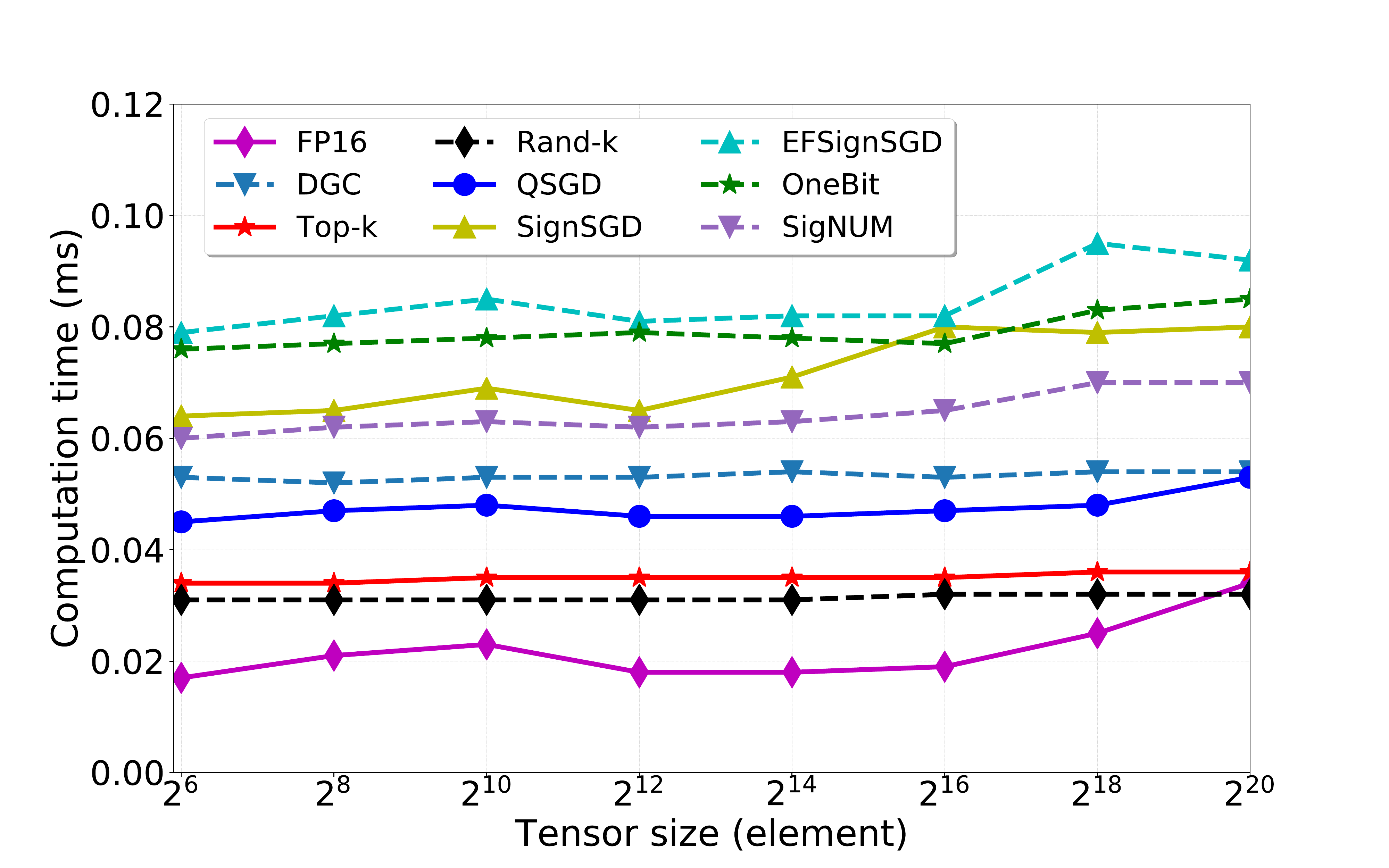}
        \caption{Decoding overhead}
    \label{fig:decomp_overhead}
    \end{subfigure} 
	\begin{subfigure}[t]{0.32\linewidth}
        \includegraphics[width=\linewidth]{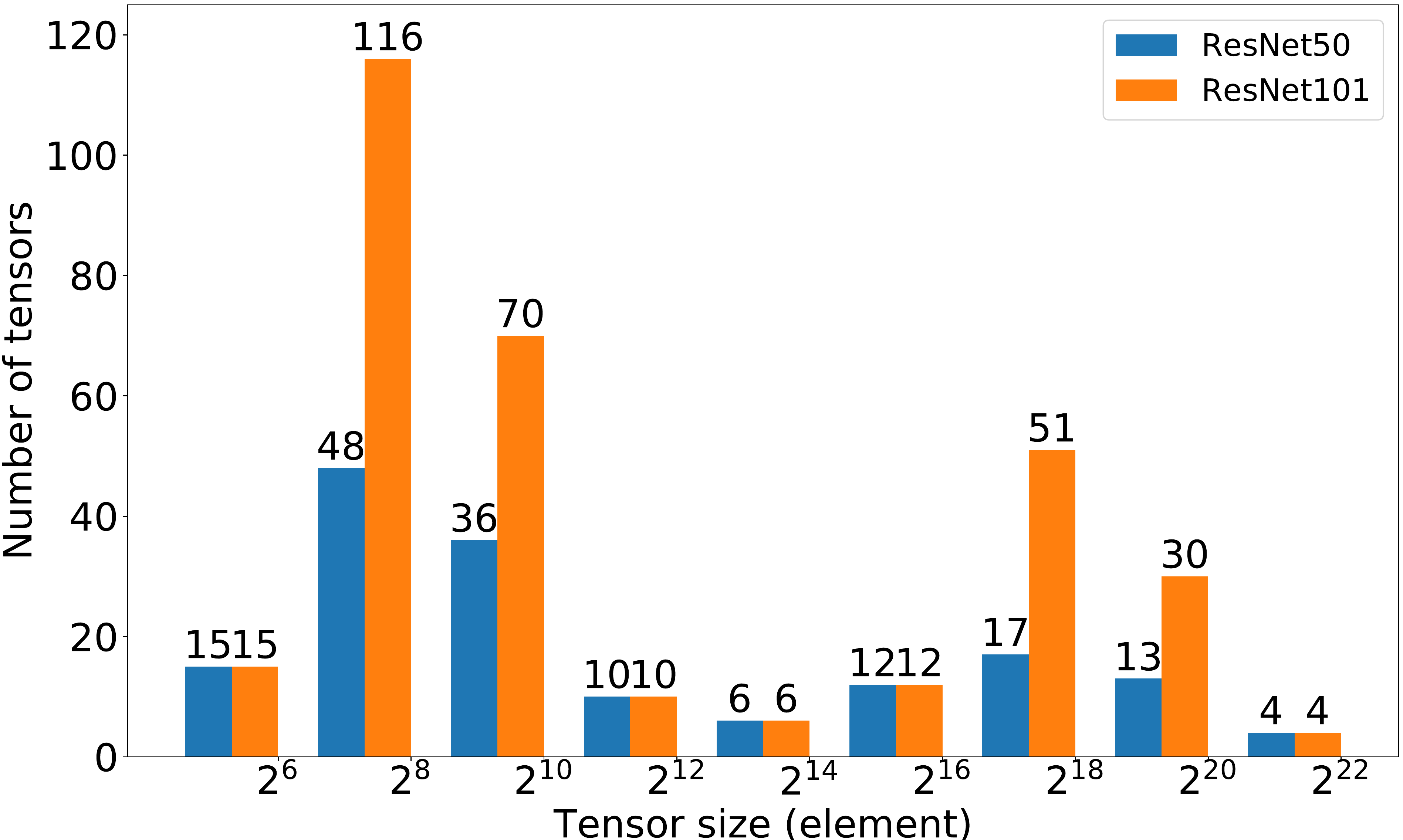}
        \caption{Number of tensors across different sizes}
        \label{fig:tensor_size}
    \end{subfigure}  
    \vskip -0.1in
    \caption{The computation overhead of each tensor with different compression algorithms. An element is an FP32 gradient. Both encoding and decoding overheads are non-negligible. The total number of tensors for synchronization in ResNet50 and ResNet101 is 161 and 314.}
    \label{fig:overhead}
    \vskip -0.1in
\end{figure*}

\begin{table}[t!]
\centering
\scriptsize
    \begin{tabular}{lcc}
    \hline
    Commnuicator             & Allreduce              & Allgather                  \\
    \hline \hline
    \multirow{2}{*}{Schemes}   & FP32             & DGC, Top-k, Rand-k, EFSignSGD \\
    ~   & FP16             & QSGD, SignSGD, OneBit, SigNUM \\
    \hline
    \multirow{2}{*}{Library} & MPI/PCIe  & \multirow{2}{*}{MPI/PCIe}           \\
    ~ & NCCL2/NVLink & ~ \\
    \hline
    \end{tabular}
    \vskip -0.1in
    \caption{The evaluated schemes and their communication settings.}
    \label{table:compress_algo}
    \vskip -0.2in
\end{table}

Figure~\ref{fig:resnet50_lw} shows the scaling factors of different compression algorithms with layer-wise compression for ResNet50 on CIFAR10~\footnote{https://github.com/kuangliu/pytorch-cifar.}.
The baselines for both NVLink and PCIe are the performance of FP32.
We observe that these compression algorithms do not scale well.
Moreover, the performance of most compression algorithms is surprisingly worse than the baseline.
Some algorithms, such as Top-k~\cite{aji2017sparse}, DGC~\cite{dgc} and OneBit~\cite{1bit}, decrease the performance by more than 30\% compared to the baseline on PCIe.

\subsection{The root cause of the poor performance}

There are two additional operations for gradient compression algorithms compared with vanilla distributed training: encoding (compress the gradients for communication) and decoding (decompress the received compressed gradients for model update).
Figure~\ref{fig:comp_overhead} and~\ref{fig:decomp_overhead} display the encoding and decoding overhead of each tensor with different compression algorithms.
Both encoding and decoding overheads are non-negligible, even for small tensor sizes. 
For instance, the encoding overhead of most compression algorithms is greater than 0.1 ms and the decoding overhead is greater than 0.03 ms, regardless of the tensor sizes.
Furthermore, many compression algorithms leverage error-feedback~\cite{1bit, strom2015scalable, wu2018error, stich2018sparsified, dgc} to preserve the accuracy, incurring another decoding operation.


DNN models tend to have a large number of tensors for gradient synchronization.
For instance, there are 161 tensors in ResNet50 and 314 tensors in ResNet101, as shown in Figure~\ref{fig:tensor_size}.
Layer-wise compression invokes encoding-decoding operations for each tensor and the costly compression overhead dramatically suppresses the communication improvement.

We take training ResNet50 on CIFAR10 with 2GPUs connected by PCIe as a concrete example to compare the overall compression overhead against the communication improvement.
In our measurement, the iteration time of single-GPU training is around 64 ms.
Without any compression, the communication overhead~\footnote{Because of the overlap between the computation and communication, the communication overhead refers to the communication time after back-propagation, unless otherwise specified.} in each iteration is about 66 ms.
Sparsification and 1-bit quantization algorithms can reduce the communication overhead to less than 5 ms thanks to the much smaller communicated data size.
However, performing gradient compression to reduce communicated data size is not free.
For instance, the overall estimated compression overheads of EFSignSGD~\cite{efsignsgd} and DGC~\cite{dgc} are around 65 ms and 120 ms, which are close to or even higher than the communication overhead without compression.
The compression overhead leads to the poor practical performance of compression algorithms and becomes their performance bottleneck.

\subsection{An opportunity and a challenge}
The layer-wise compression overhead of compression algorithms is non-negligible.
There are some fixed overheads to launch and execute kernels in CUDA~\cite{cuda_latency} and we observe that the encoding and decoding overheads remain quite stable across a wide range of tensor sizes.
For many algorithms, the compression overhead increases by less than 50\% from the tensor size of $2^6$ to $2^{20}$ elements.
This observation indicates that merging multiple tensors for one encoding-decoding operation can potentially reduce the overall compression overhead.

However, merging tensors for compression algorithms raises a new challenge: how to merge tensors to achieve the optimal performance?
There is a trade-off between the compression overhead and the communication overhead because merging tensors to reduce the compression overhead is at the cost of the overlap between the computation and communication.
For example, an extreme case of merging tensors is to apply compression algorithms to the entire model with only one encoding-decoding operation.
However, the communication cannot begin until the computation completes, resulting in the suboptimal communication overhead, as shown in Figure~\ref{fig:wfbp}.

The optimal merging strategy depends on the compression algorithms, the DNN models and the system parameters, such as the number of workers, the network bandwidth, etc. 
We propose \algo to address the challenge to realize the promised benefits of gradient compression algorithms.
\section{\algo}

In this section, we will first introduce \algo to optimize the performance of gradient compression algorithms.
We then demonstrate that \algo preserves the convergence rate of the applied compression algorithms.
We next describe how \algo automatically determines efficient merging strategies for distributed training. 
The selected merging strategy is equivalent to a schedule for applying compression operations during training.

\subsection{Overview}

\algo aims to reduce the operation overhead for compression algorithms and meanwhile overlap the computation and communication to reduce the communication overhead.

Algorithm~\ref{alg:mergecomp} describes how \algo works to optimize the performance of compression algorithms.
A model is partitioned into $y$ groups, i.e., $\mathcal{X} = \{\rm x_1, \dots, \rm x_y\}$, and $\rm x_{i, t}$ is $\rm x_i$ in the $t_{th}$ iteration.
Tensors that belong to the same group are merged and compressed together in one compression operation.
$K$ is the total number of iterations for the training.
The compression algorithm $\mathcal{C}(\cdot)$ could be either sparsification or quantization.
\algo performs an encoding-decoding operation and communication on each group in each iteration.
After compression, a set of communication schemes, such as allreduce~\cite{ring-allreduce2009, horovod}, allgather~\cite{MPIoptimization} or parameter servers~\cite{PSosdi2014}, could be used for the synchronization of different compression algorithms.
The communication is overlapped with back-propagation to reduce the communication overhead.
After communication, the compressed gradients are decompressed and aggregated to update the model.

\begin{algorithm}[t!]
   \caption{\algo for compression algorithms. The communication is overlapped with computation.}
   \label{alg:mergecomp}
   \small
\begin{algorithmic}
   \STATE {\bfseries Input:} model $\rm \mathcal{X} = \{\rm x_1, \dots, \rm x_y\}$, learning rate $\gamma$, and compression algorithm $\mathcal{C}(\cdot)$
   \FOR{$t=0$ {\bfseries to} $K-1$}
   \FOR{$i=1$ {\bfseries to} $y$}
   \STATE ${\rm g}_{i, t} \coloneqq$ stochasticGradient(${\rm x}_{i, t}$)
   \STATE ${\rm \delta}_{i, t} \coloneqq \mathcal{C}({\rm g}_{i, t})$
   \STATE $\Delta_{i, t} \coloneqq$ communicate(${\rm \delta}_{i, t}$)
   \STATE $\tilde{\rm g}_{i, t} \coloneqq$ aggregate($\mathcal{C}^{-1}(\Delta_{i, t})$)
   \STATE ${\rm x}_{i, t+1}$ = ${\rm x}_{i, t} - \gamma \tilde{\rm g}_{i, t}$
   \ENDFOR
   \ENDFOR
\end{algorithmic}
\end{algorithm}

\subsection{\algo guarantees}
We will analyze the convergence rate of distributed training with sparse or quantized communication scheduled by \algo. 
The analysis in this paper focuses on synchronous data-parallel distributed SGD.

\subsubsection{Notation and assumptions}
Let $F(\cdot)$ be the loss function we want to optimize and the studied stochastic optimization problem is 
\vskip -0.15in
\begin{equation}
\small
    \min f(x) \coloneqq \frac{1}{n} \sum\limits_{i=1}^{n}   \underbrace{\mathbb{E}_{\xi \sim \mathcal{D}_i} F_i(x; \xi)}_\text{$=: f_i(x)$},
\end{equation}
\vskip -0.15in
where $n$ is the number of workers for the training, $\mathcal{D}_i$ is a predefined distribution for the training data on worker $i$ and $F_i(x; \xi)$ is the loss computed from samples on worker $i$.

\noindent \textbf{Notation.} In the analysis of this subsection, $\left\Vert \cdot \right\Vert_2$ denotes the L2 norm of a vector or the spectral norm of a matrix; $\nabla f(\cdot)$ denotes the gradient of a function $f$; $\textbf{1}_n$ denotes the column vector in $\mathbb{R}^n$ with 1 for all elements; $f^*$ denotes the optimal solution for the stochastic optimization problem.

We employ standard assumptions of the loss function and the variance of the stochastic gradient for the analysis.
\vskip -0.1in
\begin{assume}
\label{ass:lc}
\textup{(Lipschiz continuity) $\nabla f(\cdot)^{'}$ is Lipschitz continuous with respect to the L2 norm, i.e.,}
\vskip -0.25in
\begin{equation}
\small
    \left\Vert \nabla f_i(x) - \nabla f_i(y) \right\Vert_2 \le L \left\Vert x - y \right\Vert_2 \qquad \forall x, \forall y, \forall i.
\end{equation}
\end{assume}
\vskip -0.2in
\begin{assume}
\label{ass:bv}
\textup{(Bounded variance) The variance of stochastic gradient is bounded, i.e.,}
\vskip -0.2in
\begin{equation}
\small
    \mathbb{E}_{\xi \sim \mathcal{D}_i} \left\Vert \nabla F_i(x; \xi) - \nabla f_i(x) \right\Vert_2^2 \le \sigma^2 \qquad \forall x, \forall i.
\end{equation}
\end{assume}
\vskip -0.25in
\begin{assume}
\label{ass:ub}
\textup{(Unbiasness) The stochastic gradient of $F_i(x; \xi)$ is unbiased, i.e.,}
\vskip -0.2in
\begin{equation}
\small
    \mathbb{E}_{\xi \sim \mathcal{D}_i} \nabla F_i(x; \xi) = \nabla f_i(x) \qquad \forall x, \forall i.
\end{equation}
\end{assume}
\vskip -0.2in
\begin{assume}
\label{ass:p}
\textup{Sparsification algorithms can exchange all the gradients in any $p$ consecutive iterations.}
\end{assume}
\vskip -0.1in

These assumptions are commonly employed in previous works to analyze the convergence rate of distributed SGD~\cite{jiang2018linear, QSGD, stochastic, wen2017terngrad, lian2017can}. 

\subsubsection{Sparse communication}

Previous works proved that a subset of gradients can achieve the same convergence rate as vanilla distributed SGD~\cite{stich2018sparsified, jiang2018linear, qsparse}.
Since \algo still selects important gradients in each group for sparse communication, it preserves the convergence rate as the applied sparsification algorithms.

Under Assumptions \ref{ass:lc}-\ref{ass:p}, we have the following theorem for any sparsification algorithms with \algo.
\begin{theorem}
\label{th:sparse}
    If all workers share the same training dataset, setting $\gamma=\theta \sqrt{M/K}$ where $\theta > 0$ is a constant, $M$ is the total mini-batch size on all workers, $L\gamma \le 1$ and $6np^2L^2\gamma^2 < 1$, \algo for sparsification algorithms has the convergence rate as
    \begin{equation}
    \label{eq:convergence}
    \begin{split}
        \small
        & \frac{1}{K}(\sum\limits_{t=0}^{K-1} \left\Vert \nabla f(\frac{X_t 1_n}{n}) \right\Vert_2^2) \\
        \le & \frac{4\theta^{-1}(f(x_0) - f\kstar)+2\theta L\sigma^2}{\sqrt{MK}} + \frac{2pn^2\theta^2L^2\sigma^2}{K},
    \end{split}
    \end{equation}
    if the number of iterations satisfies $K \ge 12nM\theta^2p^2L^2$.
\end{theorem}
\vskip -0.1in

Theorem~\ref{th:sparse} has a similar structure and proof to Corollary 1 in \cite{jiang2018linear}, while with different assumption of the training dataset.
In Theorem~\ref{th:sparse}, $\frac{X_j 1_n}{n}$ is the average of the gradients on all workers. 
If $K$ is large enough, the right side in (\ref{eq:convergence}) is dominated by its first term and \algo for sparsification algorithms can converge at rate $\mathcal{O}(1/\sqrt{MK})$, the same rate as vanilla SGD~\cite{stich2018sparsified}.
See the supplemental materials for the proof to Theorem~\ref{th:sparse}.

\subsubsection{Quantized communication}
Let $\mathcal{Q}(\cdot)$ denote the quantization compressor. 
The bound of expected error of $\mathcal{Q}(\cdot)$ is defined as
    $q_i = \sup \frac{\left\Vert \mathcal{Q}(\rm x_i)-\rm x_i \right\Vert_2^2}{\left\Vert \rm x_i \right\Vert_2^2}$,
where $\rm x_i$ is the gradients in the $i_{th}$ group.
We also define $q = \max\{q_i\}$ to derive the convergence rate of quantization algorithms with \algo under Assumption 1-3.

\begin{theorem}
\label{th:quantize}
    If all workers share the same training dataset and error feedback is applied, setting $\gamma=\theta \sqrt{M/K}$ where $\theta > 0$ is a constant and $(1+\frac{q}{n})L\gamma < 2$, \algo for quantization algorithms has the following convergence rate:
    \vskip -0.25in
    \begin{equation}
    \label{eq:quantize}
    \begin{split}
        \small
        & \frac{1}{K}(\sum\limits_{t=0}^{K-1} \left\Vert \nabla f(x_t) \right\Vert_2^2) \\
        \le & \frac{2\theta^{-1}(f(x_0) - f\kstar)+(1+q)\theta L\sigma^2 y}{\sqrt{MK}},
    \end{split}
    \end{equation}
    \vskip -0.15in
    if the number of iterations satisfies $K \ge M\theta^2L^2(1+\frac{q}{n})^2$.
\end{theorem}
\vskip -0.1in
Theorem~\ref{th:quantize} shows that \algo for quantization algorithms can also converge at rate $\mathcal{O}(1/\sqrt{MK})$.
See the supplemental materials for the proof to Theorem~\ref{th:quantize}.

\subsection{Searching for the optimal model partition}
We will explore how to determine the optimal partitioning strategy for distributed training.
We formulate the model partition problem as an optimization problem. 

\textbf{Notation.}  
Let $N$ denote the number of tensors of a model and $A$  the computation time of each iteration.
Let $y$ be the number of partitioned groups, $x_i$ the size of the $i_{th}$ group (i.e., $x_i = |\rm x_i|$), and $X_y = \{x_1, \dots, x_y\}$ a partition of the model. 
Recall that tensors belonging to the same group are merged for compression.
$h(x_i)$ is the compression time of the $i_{th}$ group, $g(x_i)$ is its communication time, and $p(x_i)$ is the overlap time between the computation and communication time.

The goal of searching for the optimal partition is to minimize the iteration time for compression algorithms.
Without WFBP, the iteration time is the summation of the computation time, the compression time, and the communication time; while with WFBP, the overlap time helps reduce the communication overhead.
Therefore, the optimization objective of the model partition problem is to 
\vskip -0.3in
\begin{equation}
    \small
    \min \limits_{y \in [1,N]} 
    \min \limits_{X_y \in \mathcal{X}_y} 
    F(X_y) \coloneqq A + \sum\limits_{i=1}^{y} h(x_i) + \sum\limits_{i=1}^{y} g(x_i) - \sum\limits_{i=1}^{y}p(x_i),
\end{equation}
where $\mathcal{X}_y$ is the set of all possible partitions with $y$ groups.

\begin{lemma}
    \label{lemma:space}
    \textup{The size of the search space for the model partition problem is $2^{N-1}$.}
\end{lemma}

\vskip -0.1in
Lemma~\ref{lemma:space} indicates that the general form of the model partition problem cannot be solved in polynomial time.
In practice, the compression time and the communication time both increases with the group size.
Therefore, we make the following assumption for the model partition problem.

\begin{assume}
\label{assume:linear}
\textup{(Linear overhead) $h(x_i) = B_h + \gamma_h x_i$ and $g(x_i) = B_g + \gamma_g x_i$, where $B_h$ and $B_g$ are the latency (or startup time) for the compression and communication time; $\gamma_h$ and $\gamma_g$ are the overheads for each unit.}
\end{assume}

Under Assumption~\ref{assume:linear}, we have the following lemma.
\begin{lemma}
\label{lemma:y}
\textup{Given a particular $y$, all possible partitions have the same compression time and communication time; and they both increase with the value of $y$.}
\vskip -0.6in
\end{lemma}

\vskip -0.1in

\begin{algorithm}[t!]
   \caption{A heuristic algorithm for model partition}
   \label{alg:partition}
   \small
\begin{algorithmic}
    \STATE {\bfseries Input:} $Y \in [2, N]$, $\alpha \in (0, 1)$
    \STATE $F_{min}(1) = F(X_1)$
    \FOR{$y=2$ {\bfseries to} $Y$}
    \STATE $X_y\kstar = \argmin \limits_{X_y \in \mathcal{X}_y} F(X_y) $
       \STATE $F_{min}(y) = F(X_y\kstar)$
       \IF{$F_{min}(y-1) < F_{min}(y)$}
            \STATE \textup{\textbf{return}} $X_{y-1}\kstar$
       \ELSIF{$F_{min}(y-1) - F_{min}(y) < \alpha F_{min}(y-1)$}
            \STATE \textup{\textbf{return}} $X_{y}\kstar$
        \ENDIF
   \ENDFOR
   \STATE \textup{\textbf{return}} $X_{y}\kstar$
\end{algorithmic}
\end{algorithm}

There are two empirical observations for the model partition problem: 1) increasing the number of partition groups helps increase the overlap time, but it also increases the overhead of compression and communication time according to Lemma~\ref{lemma:y}; and 2) the marginal benefit of the overlap time is diminishing with the increasing number of groups.
Based on these two observations, we propose a heuristic algorithm to approximately solve the model partition problem in polynomial time, as shown in Algorithm~\ref{alg:partition}.

\algo searches for an efficient model partition with Algorithm~\ref{alg:partition} at the beginning of training.
$Y$ and $\alpha$ are two parameters to narrow down the search space. 
For any $y \in [2, Y]$, the algorithm searches for the optimal partition $X_y\kstar$ (refer to the proof to Theorem~\ref{th:search} for the details) and then compares the iteration time with $X_{y-1}\kstar$.
If its performance is worse than $X_{y-1}\kstar$ or the marginal benefit is less than $\alpha F_{min}(y-1)$, the algorithm terminates the search and the best partition strategy discovered by Algorithm 2 is used for the remaining training iterations.

\begin{theorem}
    \label{th:search}
    The time complexity of Algorithm~\ref{alg:partition} to solve the model partition problem is $O(N^{Y-2}\log N)$.
\end{theorem}
\vskip -0.1in

See the supplemental materials for the proof to Theorem~\ref{th:search}.
We experimentally observe that Algorithm~\ref{alg:partition} works very well in practice to achieve good scalability for distributed training.
Furthermore, it makes no assumption of model architectures or system parameters (e.g., the batch size, the number of GPUs, and network bandwidth capacities) to search for an efficient model partition.
\section{Experiments}

In this section, we will first show the performance improvement of \algo for various compression algorithms over both PCIe and NVLink. 
We then use end-to-end experiments to demonstrate that \algo can preserve the accuracy of the applied compression algorithms.

\begin{figure*}[ht!]
    \centering
    \begin{subfigure}[t]{0.95\linewidth}
    \centering
	\includegraphics[width=0.5\linewidth]{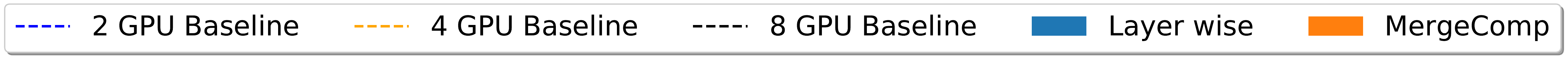}
	\label{fig:speed_bar}
    \end{subfigure}
    \begin{subfigure}[t]{0.33\linewidth}
	\includegraphics[width=\linewidth]{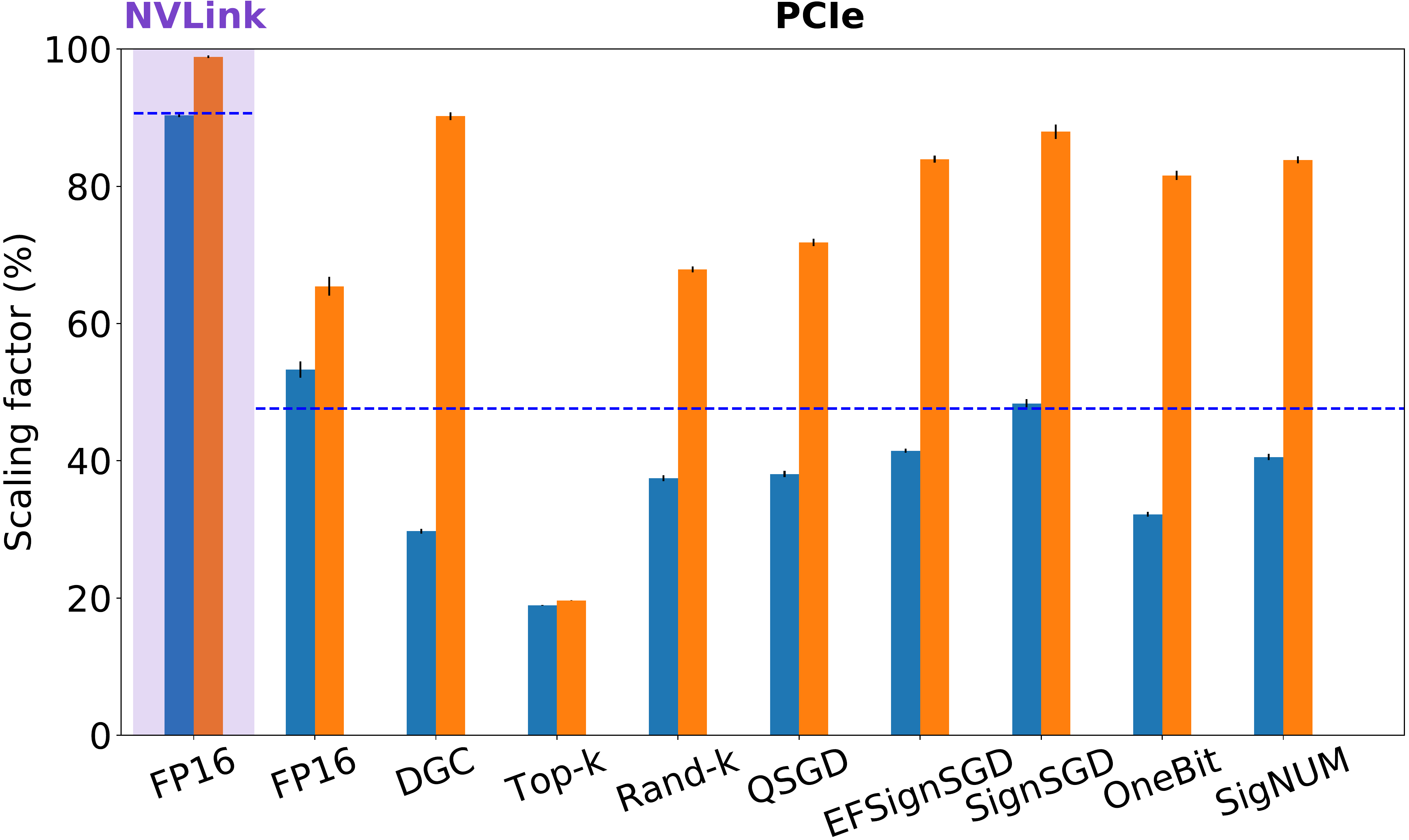}
	\caption{2 GPUs}
	\label{fig:resnet50_2GPU}
    \end{subfigure}    
	\begin{subfigure}[t]{0.33\linewidth}
	\includegraphics[width=\linewidth]{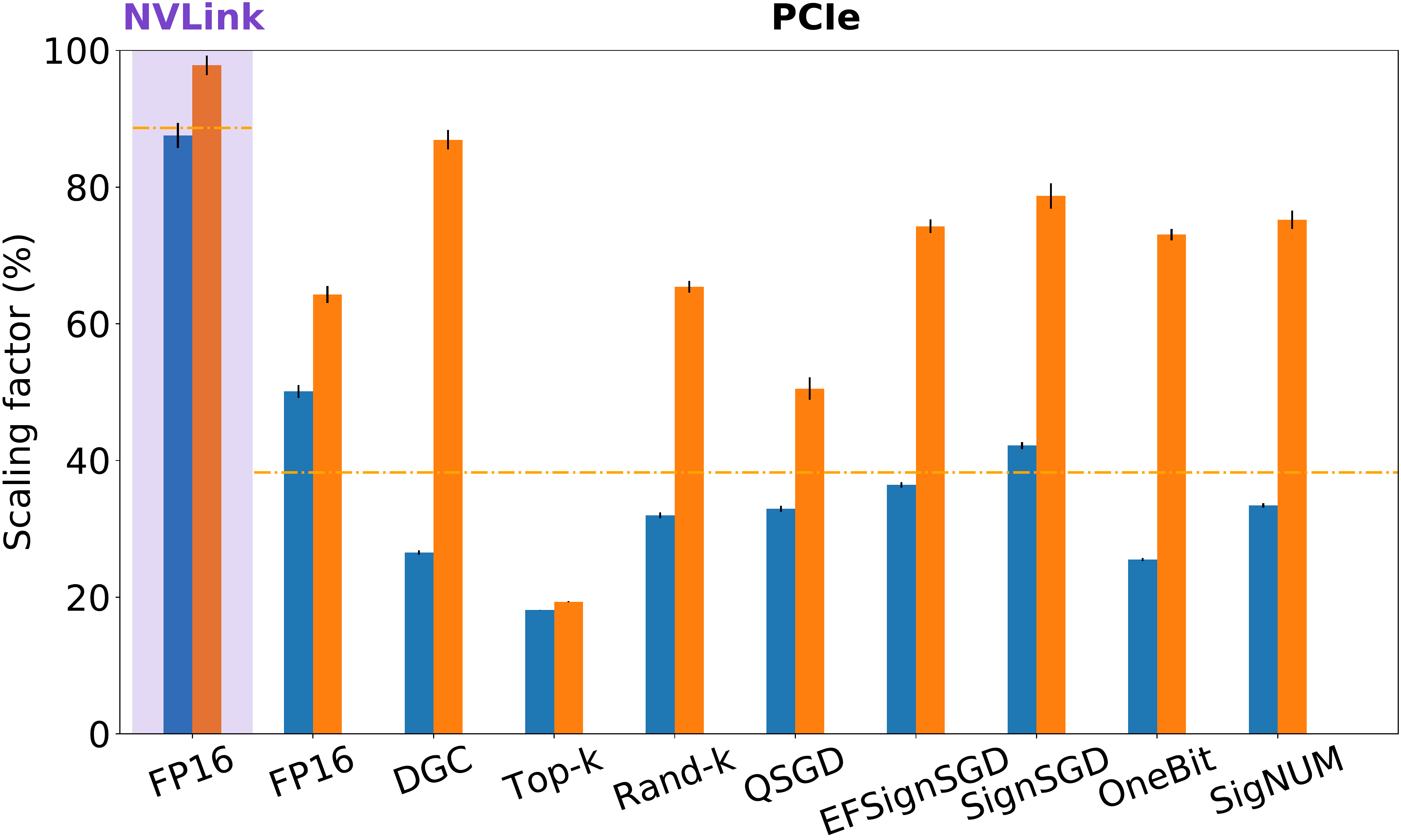}
	\caption{4 GPUs}
	\label{fig:resnet50_4GPU}
    \end{subfigure} 
    \begin{subfigure}[t]{0.33\linewidth}
	\includegraphics[width=\linewidth]{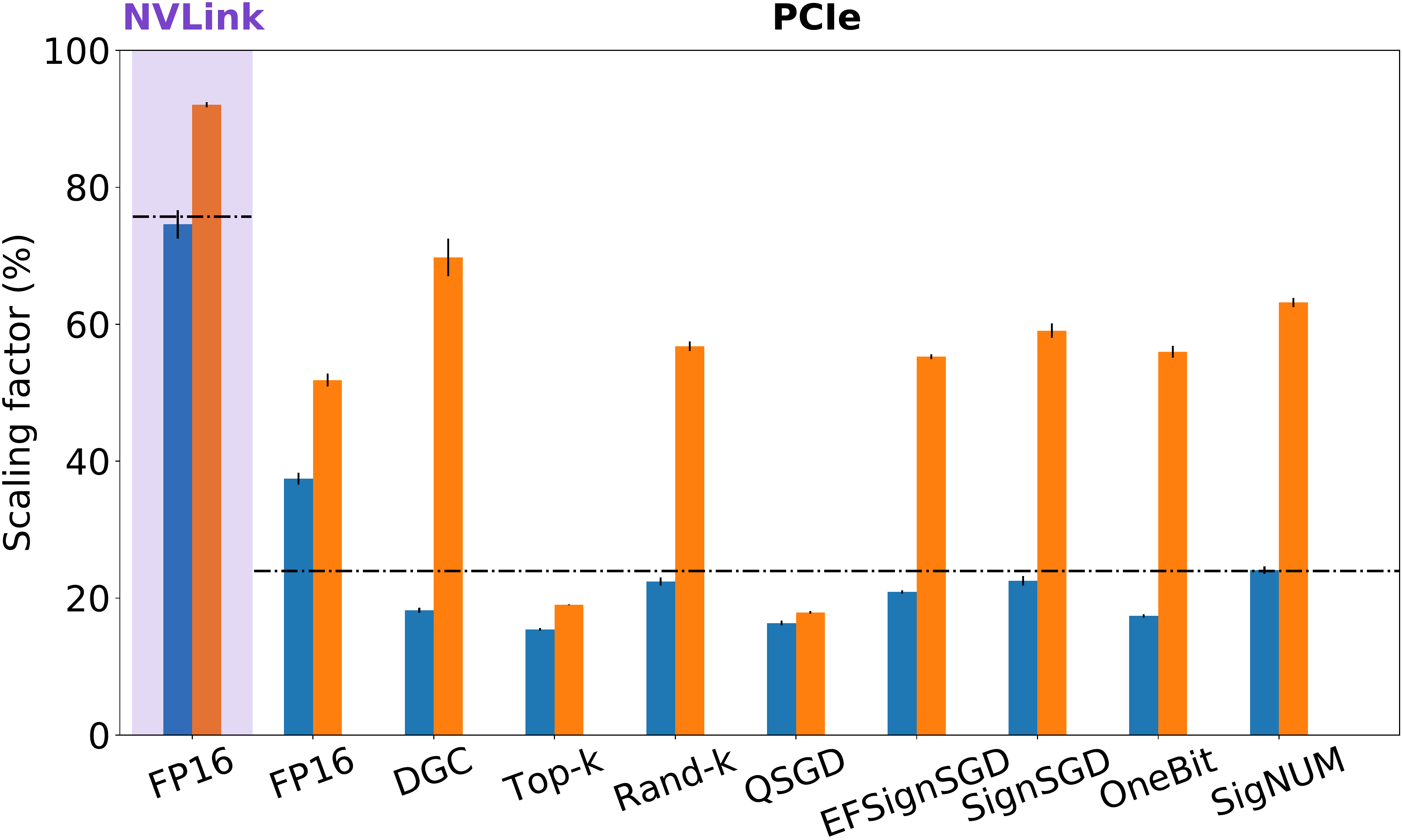}
	\caption{8 GPUs}
	\label{fig:resnet50_8GPU}
    \end{subfigure} 
    \vskip -0.1in
    \caption{The performance of ResNet50 on CIFAR10. The scaling factor of \algo is up to 2.91$\times$ and 3.83$\times$ (DGC with 8 GPUs) higher than that of the baseline and layer-wise compression.}
    \label{fig:resnet50_sf}
    \vskip -0.1in
\end{figure*}

\begin{figure*}[ht!]
    \centering
    \begin{subfigure}[t]{0.33\linewidth}
	\includegraphics[width=\linewidth]{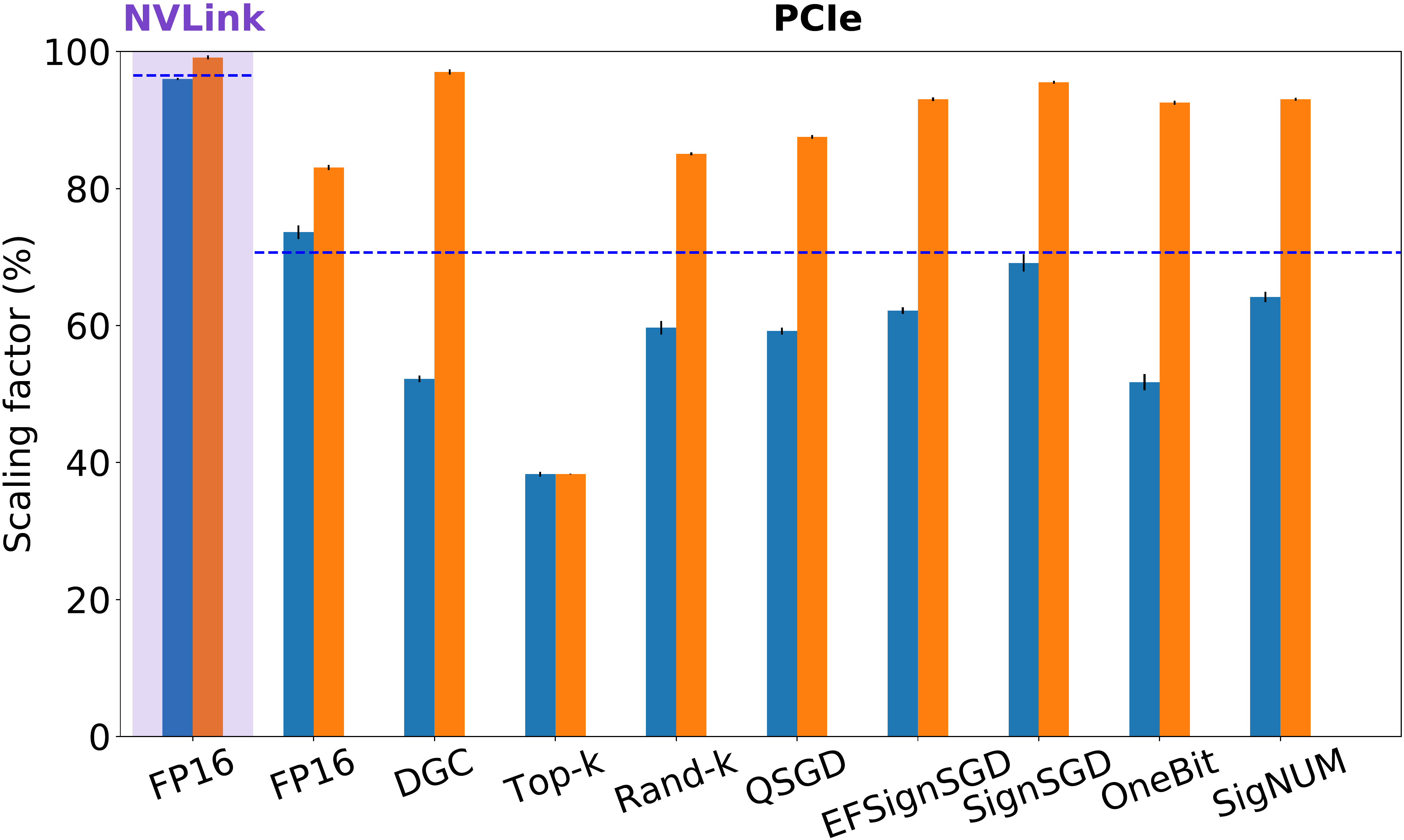}
	\caption{2 GPUs}
	\label{fig:resnet101_2GPU}
    \end{subfigure}    
	\begin{subfigure}[t]{0.33\linewidth}
	\includegraphics[width=\linewidth]{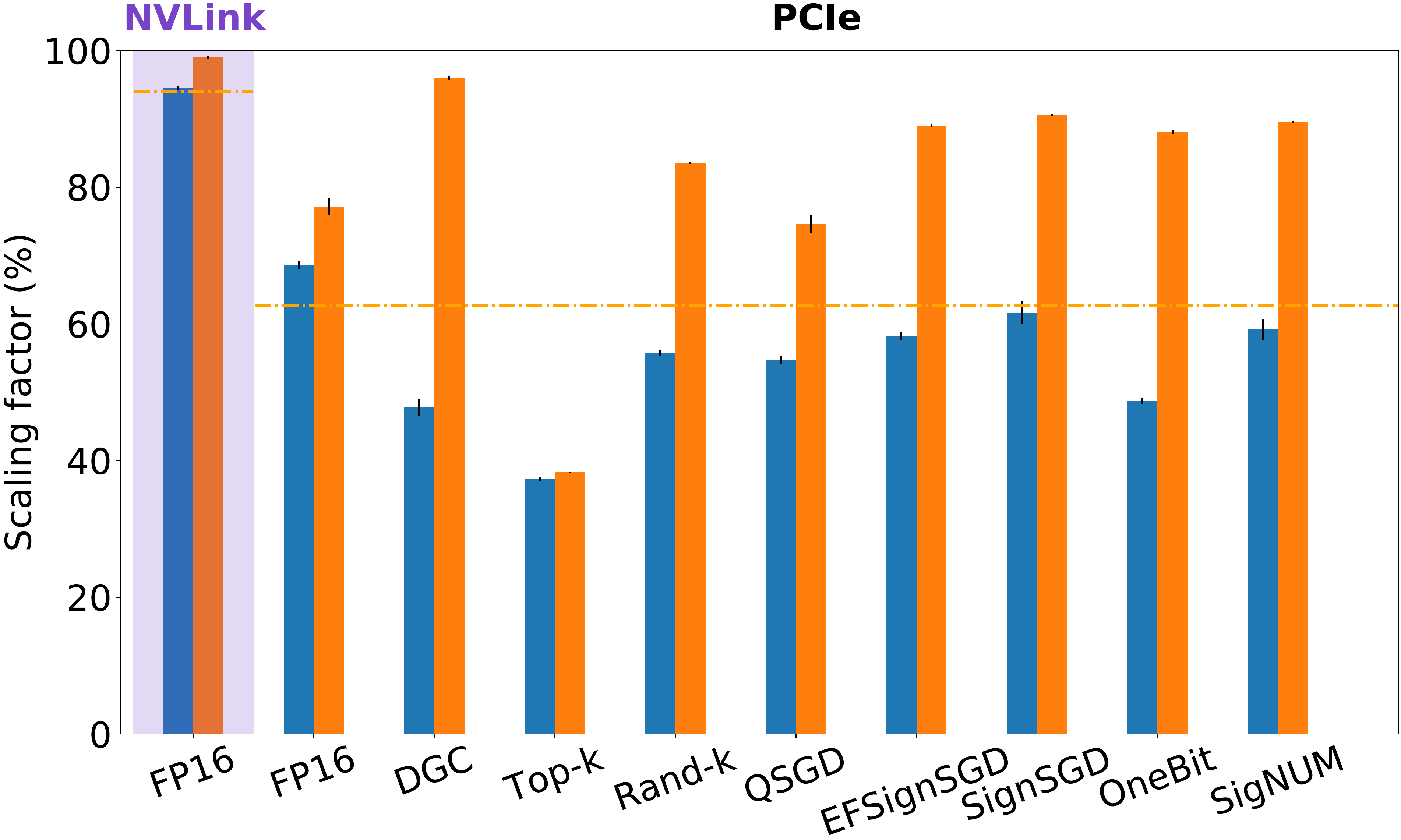}
	\caption{4 GPUs}
	\label{fig:resnet101_4GPU}
    \end{subfigure} 
    \begin{subfigure}[t]{0.33\linewidth}
	\includegraphics[width=\linewidth]{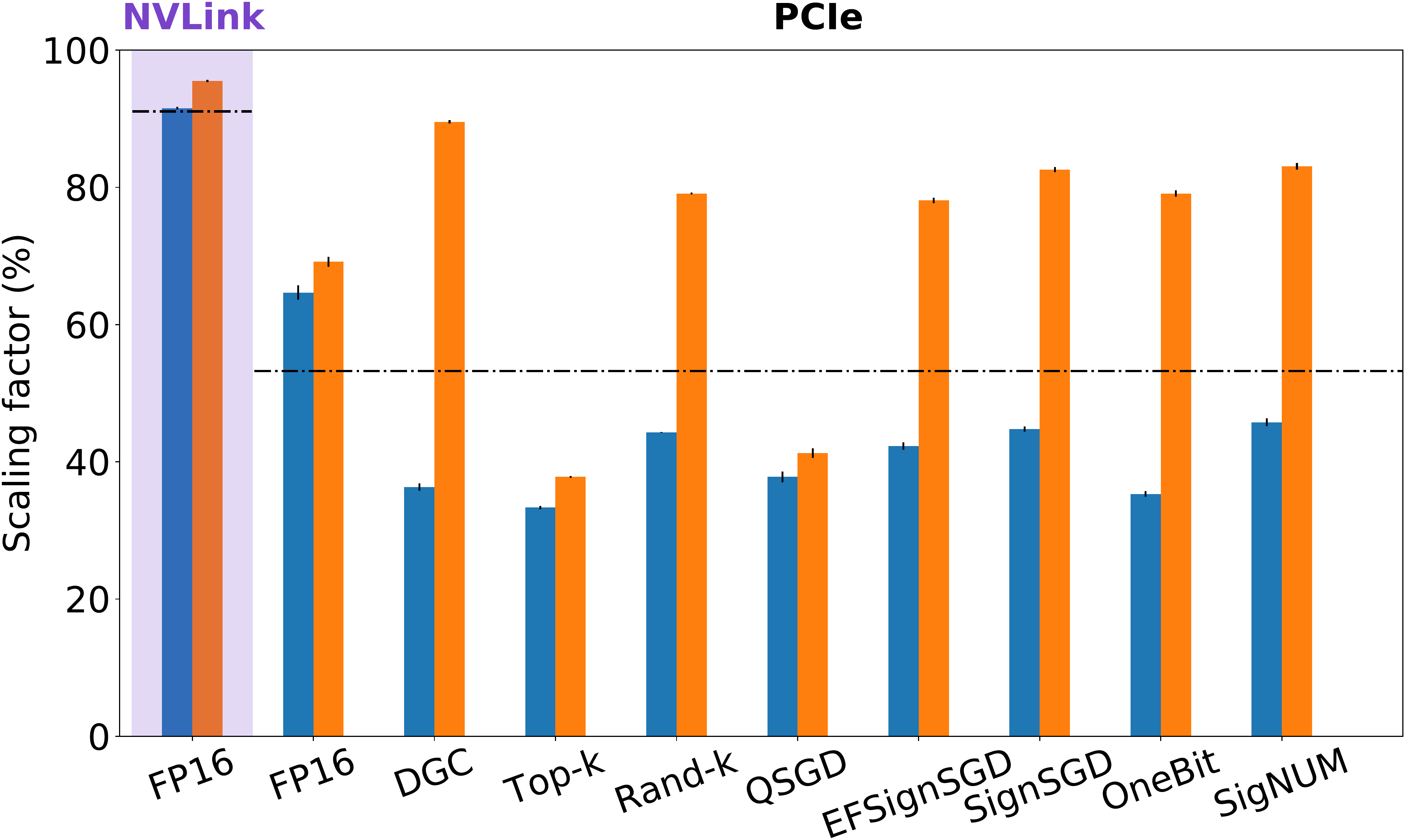}
	\caption{8 GPUs}
	\label{fig:resnet101_8GPU}
    \end{subfigure} 
    \vskip -0.1in
    \caption{The performance of ResNet101 on ImageNet. The scaling factor of \algo is up to 1.68$\times$ and 2.46$\times$ (DGC with 8 GPUs) higher than that of the baseline and layer-wise compression. Legend in Figure~\ref{fig:resnet50_sf}.}
    \label{fig:resnet101_sf}
    \vskip -0.1in
\end{figure*}

\begin{figure*}[ht!]
    \centering
    \begin{subfigure}[t]{0.33\linewidth}
	\includegraphics[width=\linewidth]{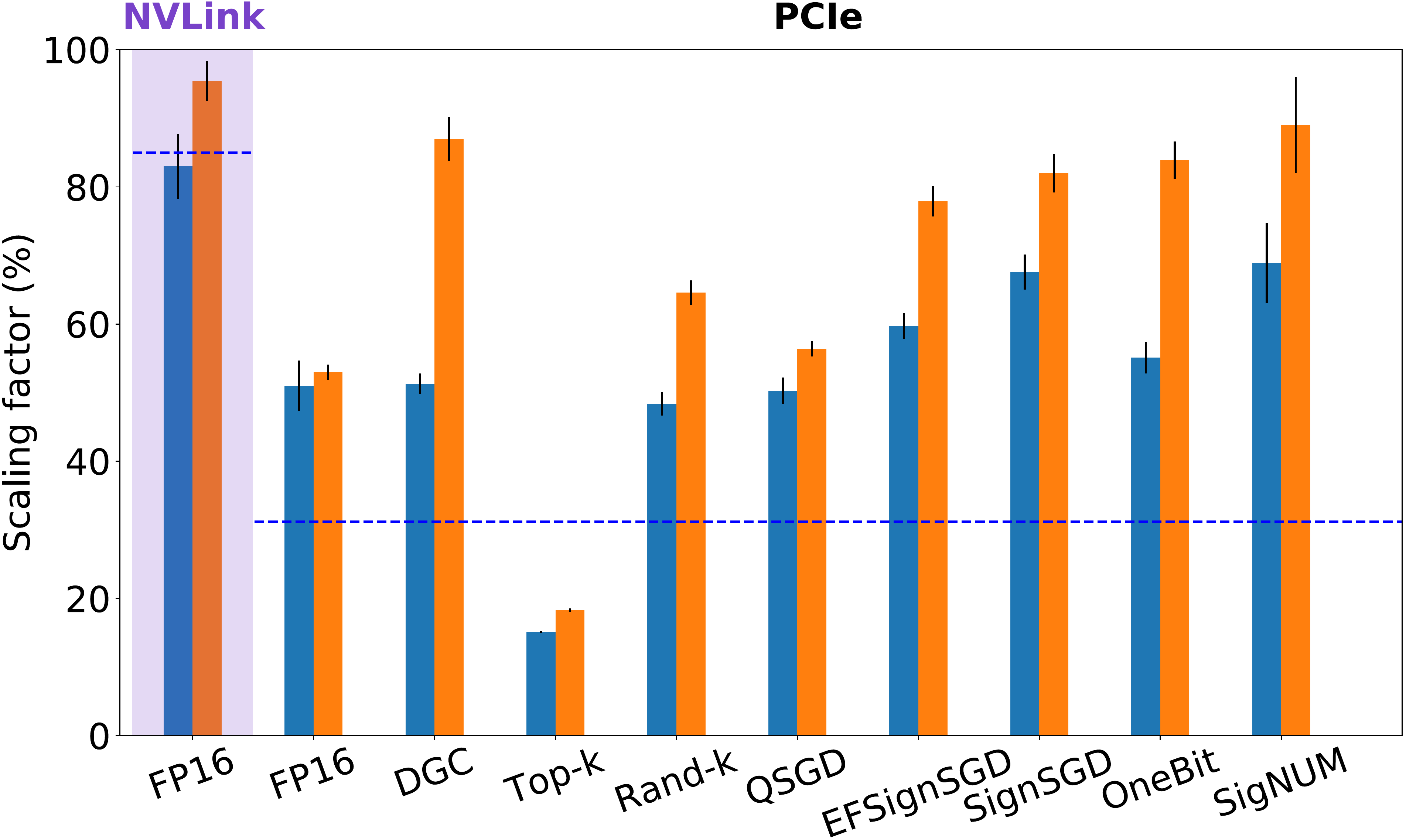}
	\caption{2 GPUs}
	\label{fig:maskrcnn_2GPU}
    \end{subfigure}    
	\begin{subfigure}[t]{0.33\linewidth}
	\includegraphics[width=\linewidth]{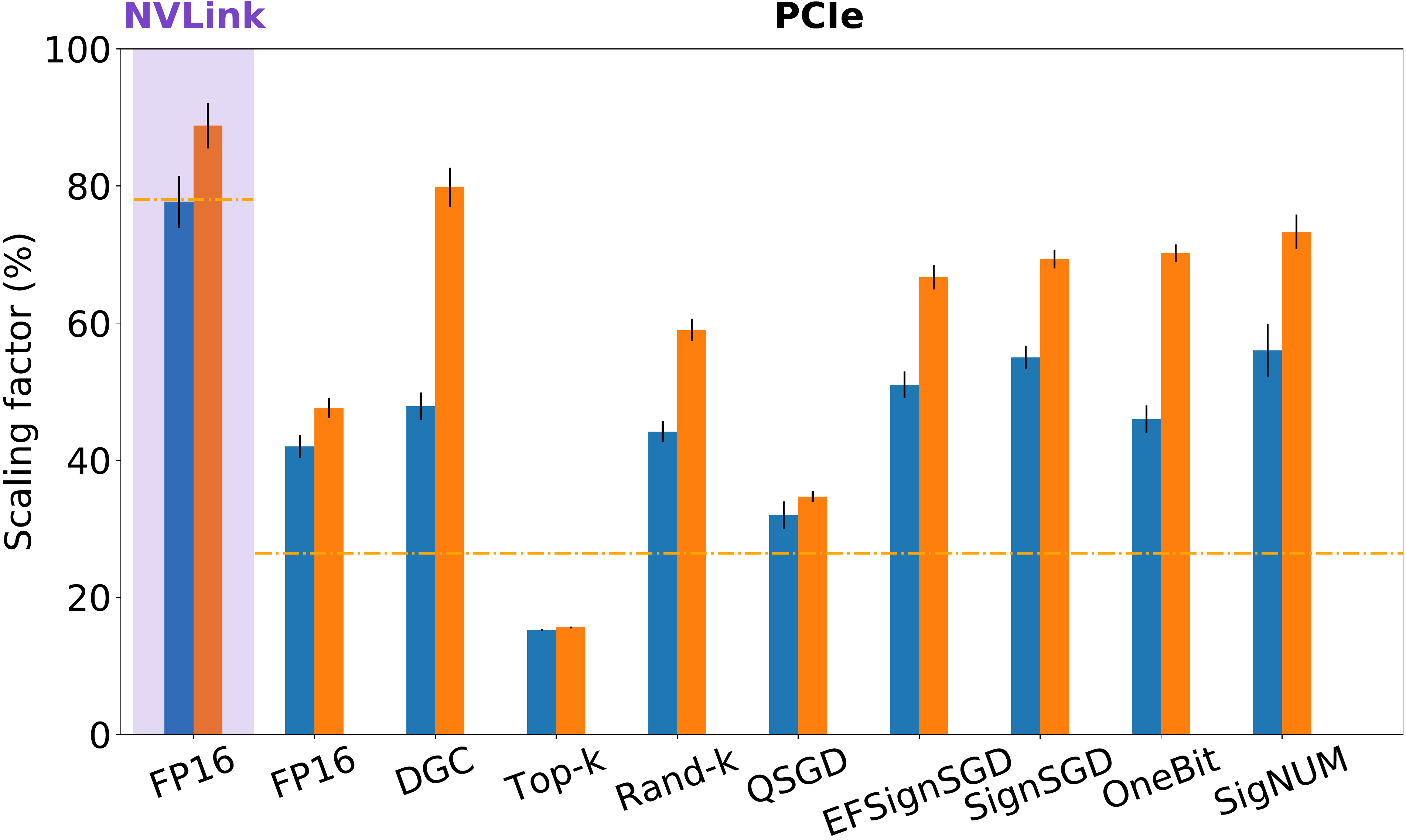}
	\caption{4 GPUs}
	\label{fig:maskrcnn_4GPU}
    \end{subfigure} 
    \begin{subfigure}[t]{0.33\linewidth}
	\includegraphics[width=\linewidth]{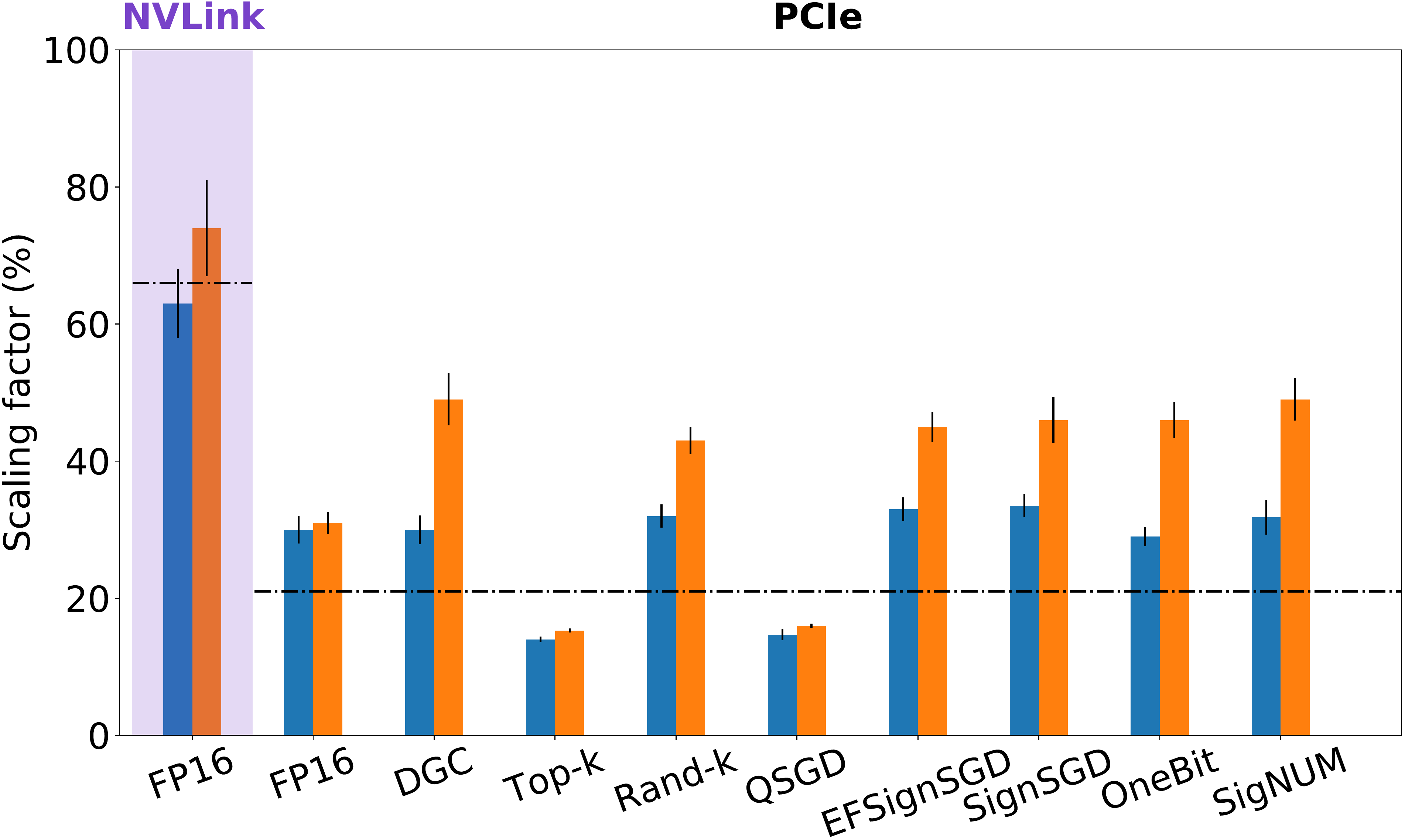}
	\caption{8 GPUs}
	\label{fig:maskrcnn_8GPU}
    \end{subfigure} 
    \vskip -0.1in
    \caption{The performance of Mask R-CNN on COCO. The scaling factor of \algo is up to 2.33$\times$ and 1.66$\times$ (DGC with 8 GPUs) higher than that of the baseline and layer-wise compression. Legend in Figure~\ref{fig:resnet50_sf}.}
    \label{fig:maskrcnn_sf}
    \vskip -0.15in
\end{figure*}

\textbf{Setup.} 
The setup is the same as that described in Section~\ref{sec:measurement}.

\textbf{Workloads.}
We validate the performance of \algo on two types of machine learning tasks: image classification and image segmentation.
The models include ResNet50 and ResNet101~\cite{ResNet-50} on CIFAR10~\cite{cifar10} and ImageNet~\cite{imagenet}; Mask R-CNN~\cite{maskrcnn} on COCO~\cite{coco}.
These three models are widely used as standard benchmarks to evaluation the scalability of distributed training.
The default batch size for image classification is 64 and image segmentation is 1.

\textbf{Methods.} We benchmark the tasks with \algo and set $Y=2, 3, 4$ for the evaluation. 
We compare them against FP32 (baseline) and layer-wise compression. 
The gradient sparsity of sparsification algorithms is 99\% and each FP32 element is mapped to 8 bits in QSGD~\cite{QSGD}.

\textbf{Metrics.} We use the scaling factor and Top-1 accuracy as evaluation metrics. 
The results for scaling factors are reported with the average of 20 runs. 
We also report the standard deviation using the error bar because the training speed varies at times.

\subsection{Training speed improvement}

We apply \algo to nine popular compression algorithms for three DNN models.
Figures~\ref{fig:resnet50_sf}-\ref{fig:maskrcnn_sf} present the performance of various compression algorithms with $Y=2$. 
We omit the results of \algo with $Y=3$ and $Y=4$ since they have very similar performance to that with $Y=2$. 
We evaluate the choice of $Y$ in Section~\ref{sec:partition_eval}.

\algo can significantly improve the performance of both sparsification and quantization algorithms.
For instance, on PCIe, the scaling factor of \algo with DGC for ResNet50 on CIFAR10 is up to 2.91$\times$ and 3.83$\times$ higher than that of the baseline and layer-wise compression.
Similarly, \algo improves the scaling factor of ResNet101 on ImageNet by 1.68$\times$ and 2.46$\times$ compared to the baseline and layer-wise compression.
There is no obvious improvement for Top-k because its performance bottleneck is still the compression overhead, i.e., the time-consuming top-k() operation.

\algo can help distributed training achieve near-linear scalability on NVLink.
With FP32, the scaling factor of ResNet50 on CIFAR10 with 8 GPUs is about 75\%.
Converting the gradients from FP32 to FP16 for communication decreases the performance because of the compression overhead, but applying \algo to FP16 improves the scaling factors to 92\%.
Similarly, \algo achieves the scaling factors of 99\% and 96\% for ResNet101 on ImageNet with 4 GPUs and 8 GPUs, respectively.

For 1 bit quantization algorithms, e.g., EFSignSGD~\cite{efsignsgd}, SignSGD~\cite{signsgd}, OneBit~\cite{1bit} and SigNUM~\cite{signum}, the scaling factor of \algo for ResNet50 is up to 2.60$\times$ and 3.18$\times$ higher than that of the baseline and layer-wise compression.
The corresponding improvements for ResNet101 are 1.56$\times$ and 2.24$\times$, respectively.

Layer-wise compression for Mask R-CNN has better performance than the baseline over PCIe because it has relatively few tensors so the layer-wise compression overhead is not too excessive.
\algo still outperforms layer-wise compression by up to 1.66$\times$ on PCIe and 1.1$\times$ on NVLink.

\subsection{The model partition algorithm}
\label{sec:partition_eval}
We evaluate the effectiveness of the proposed heuristic algorithm, which searches for the near-optimal partitioning strategy for distributed training.
The evaluated model is ResNet101 on ImageNet. 
\algo is applied to three representative compression algorithms: FP16, DGC and EFSignSGD. 

Table~\ref{table:comp_y} shows the performance of \algo with different values of $Y$. 
The results are normalized against the performance of \algo with $Y=1$.
It shows that partitioning the model into more than one groups could improve the scalability and the achieved improvement of \algo increases with the number of GPUs.
It is because the communication overhead increases with the number of GPUs and \algo can search for the near-optimal partition to optimize the communication overhead. 

The performance of \algo with $Y=3$ and $Y=4$ is very close to that with $Y=2$~\footnote{We omit the results of \algo with $Y=4$ since its performance is similar to that with $Y=2$ and $Y=3$.}.
The observation is that it is a good choice to set $Y=2$ for \algo because the marginal benefit of a larger $Y$ is negligible.
Moreover, because the time complexity of Algorithm~\ref{alg:partition} is $O(N^{Y-2}\log N)$, a larger $Y$ requires many more iterations to search for an efficient model partition than $Y=2$, which needs less than 50 iterations in our evaluation.

We also compare the performance of \algo with a naive partition strategy, which evenly partitions the number of tensors to each group.
As shown in Table~\ref{table:comp}, the performance of \algo is up to 5.5\% higher than that of the naive partition with $Y=2$.

\begin{table}[t!]
\scriptsize
\centering
    \begin{tabular}{lllllll}
    \hline
    \multirow{2}{*}{Compressor} & \multicolumn{3}{c}{$Y=2$}  & \multicolumn{3}{c}{$Y=3$}  \\
    ~          & 2GPUs          & 4GPUs & 8GPUs & 2GPUs    & 4GPUs & 8GPUs \\
    \hline \hline
    FP16       & 1.16$\times$      & 1.18$\times$   & 1.23$\times$   & 1.16$\times$      & 1.18$\times$   & 1.23$\times$ \\ 
    DGC        & 1.04$\times$       & 1.06$\times$    & 1.06$\times$   & 1.04$\times$       & 1.06$\times$    & 1.06$\times$ \\
    EFSignSGD  & 1.04$\times$       & 1.05$\times$    & 1.13$\times$   & 1.04$\times$       & 1.04$\times$    & 1.13$\times$  \\
    \hline
    \end{tabular}
    \vskip -0.1in
    \caption {The performance of \algo with different number of partition groups $Y$. The numbers are normalized against the performance of \algo with $Y=1$.}
    \label{table:comp_y}
\end{table}

\begin{table}[t!]
\scriptsize
\centering
    \begin{tabular}{lllllll}
    \hline
    Compressor & 2 GPUs    & 4 GPUs & 8 GPUs \\
    \hline \hline
    FP16       & 5.5\%  & 5.4\%   & 5.1\%  \\
    DGC        & 2.0\%  & 1.9\%    & 1.9\% \\
    EFSignSGD  & 3.4\%  & 3.3\%   & 3.1\%  \\
    \hline
    \end{tabular}
    \vskip -0.1in
    \caption {The performance improvement of \algo compared against the naive partition with $Y=2$.}
    \label{table:comp}
\end{table}

\subsection{End-to-end experiments}

\begin{figure}[t!]
    \centering
    \begin{subfigure}[t]{0.49\linewidth}
	\includegraphics[width=\linewidth]{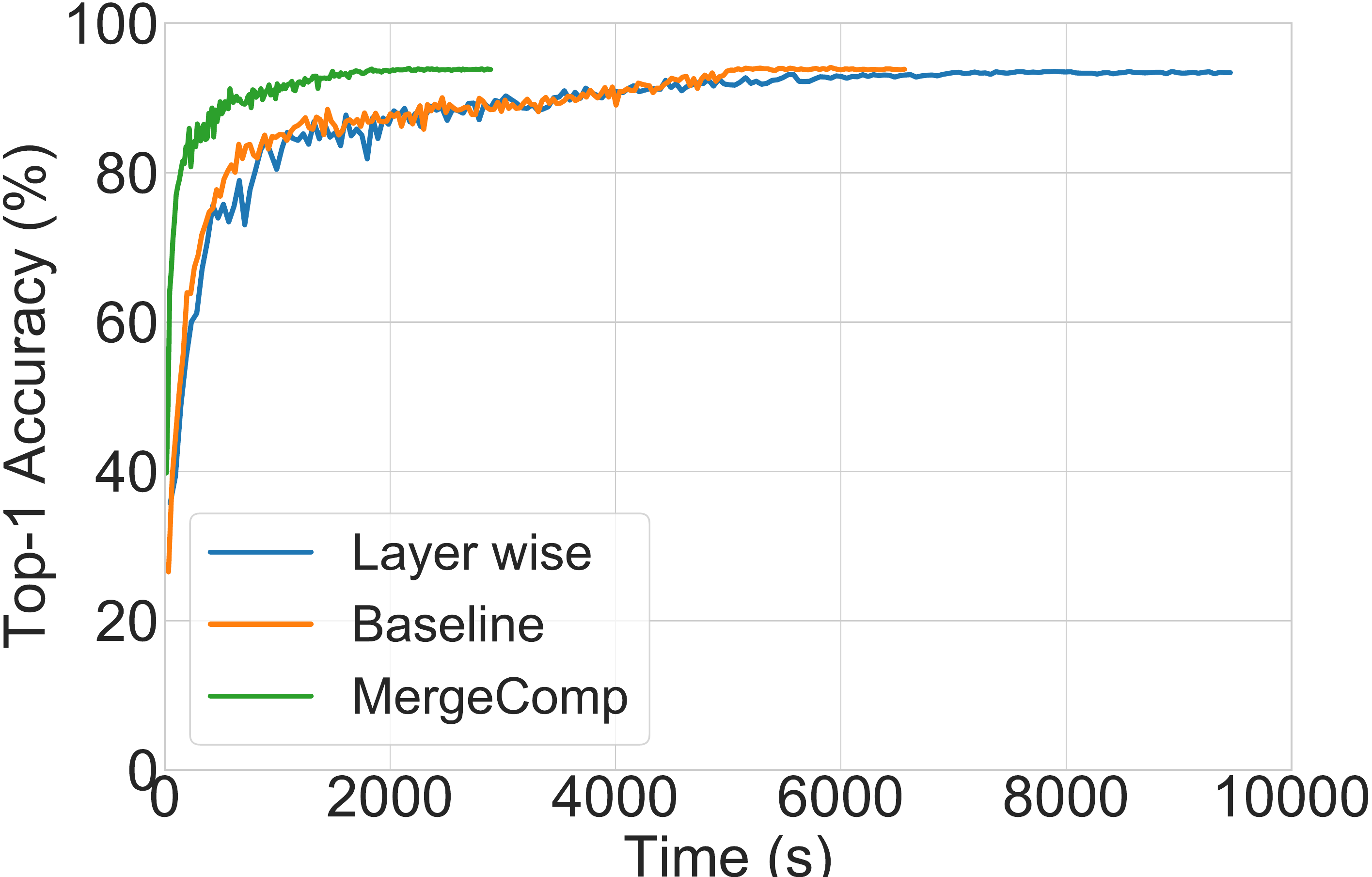}
	\label{fig:cifar10_dgc_time}
    \end{subfigure}    
	\begin{subfigure}[t]{0.49\linewidth}
	\includegraphics[width=\linewidth]{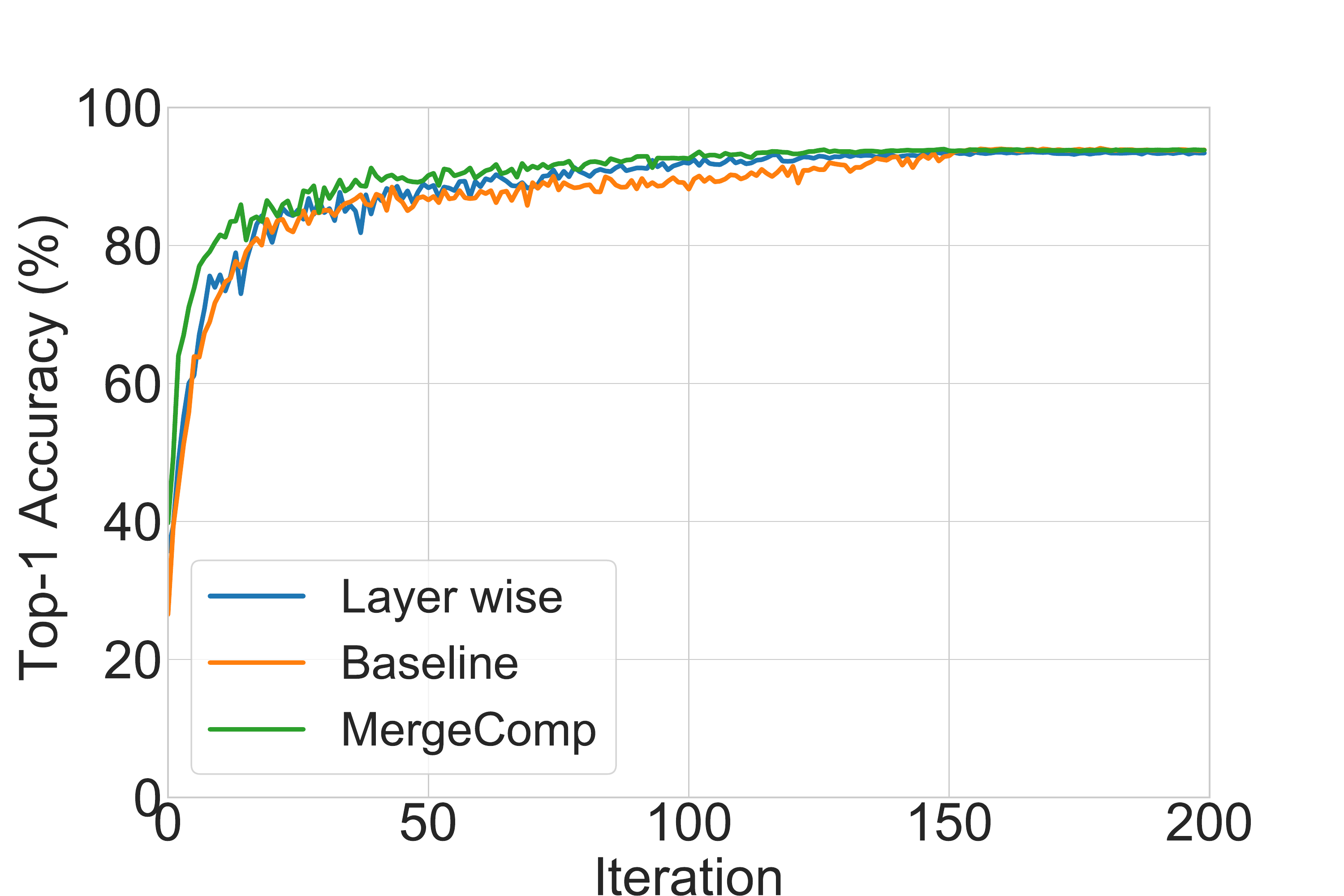}
	\label{fig:cifar10_dgc_iter}
    \end{subfigure} 
    \vskip -0.2in
    \caption{The performance of \algo for ResNet50 on CIFAR10. The applied compression algorithm is DGC. The time required for convergence with \algo is 2.27$\times$ and 3.28$\times$ lower than that of the baseline and layer-wise compression.}
    \label{fig:cifar10_accuracy}
    \vskip -0.1in
\end{figure}

\begin{figure}[t!]
    \centering
    \begin{subfigure}[t]{0.49\linewidth}
	\includegraphics[width=\linewidth]{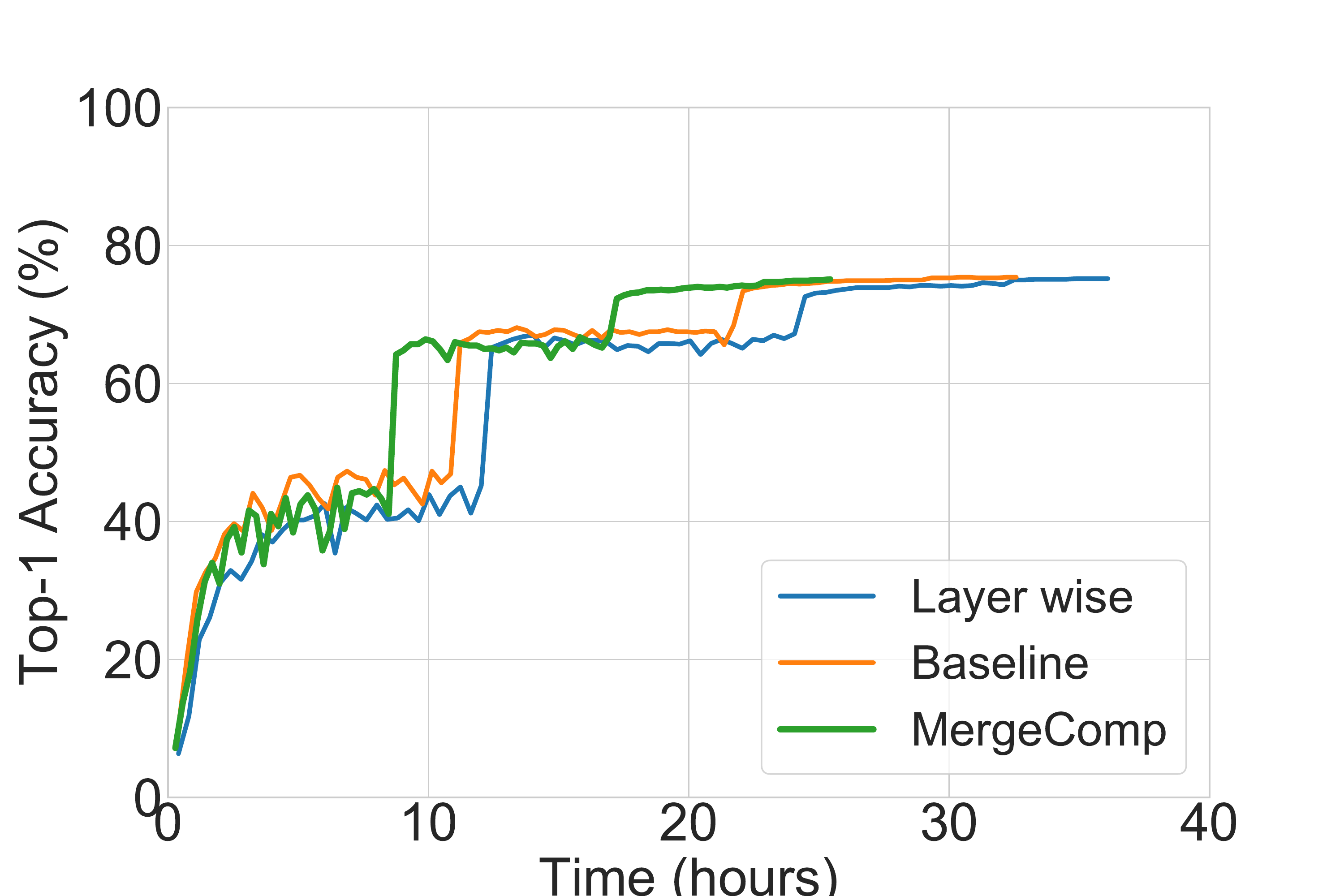}
	\label{fig:imagenet_time}
    \end{subfigure}    
	\begin{subfigure}[t]{0.49\linewidth}
	\includegraphics[width=\linewidth]{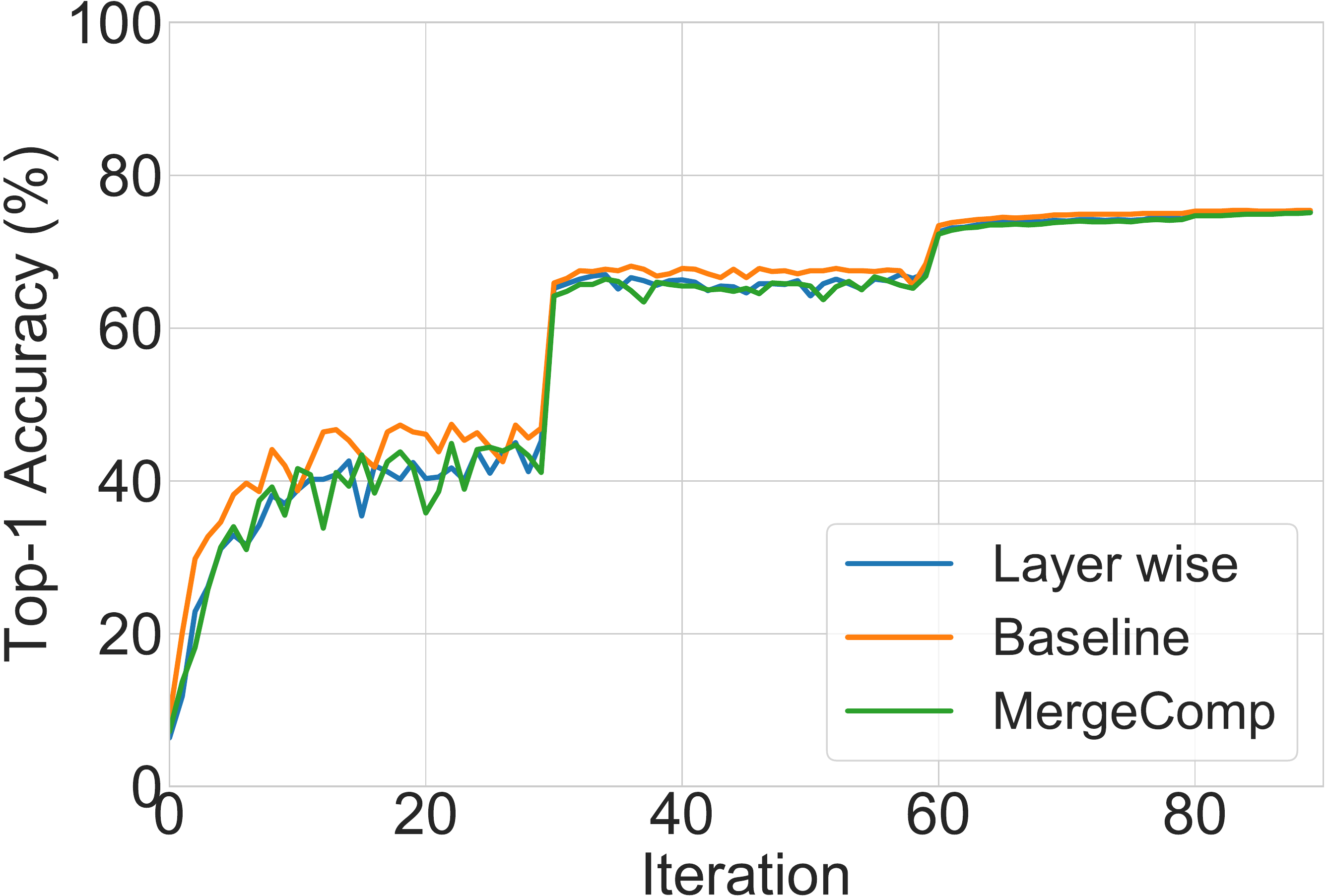}
	\label{fig:imagenet_iter}
    \end{subfigure} 
    \vskip -0.2in
    \caption{The performance of \algo for ResNet50 on ImageNet. The applied compression algorithm is EFSignSGD.}
    \label{fig:imagenet_accuracy}
    \vskip -0.2in
\end{figure}

We apply \algo to EFSignSGD and DGC to show its performance for end-to-end training with 4 GPUs connected by PCIe.
Figure~\ref{fig:cifar10_accuracy} illustrates the time and iteration wise comparisons of ResNet50 on CIFAR10.
The time required for convergence with \algo is 2.27$\times$ and 3.28$\times$ lower than that of the baseline and layer-wise compression.

Similarly, Figure~\ref{fig:imagenet_accuracy} illustrates the performance comparison for ResNet50 on ImageNet.
The time required for convergence with \algo is 1.28$\times$ and 1.42$\times$ lower than that of the baseline and layer-wise compression; and the scaling factor of \algo achieves around 90\%.

\begin{table}[t!]
\centering
\scriptsize
\tabcolsep=0.11cm
    \begin{tabular}{lllc}
    \hline
    Compressor & Datasets & Methods      & Top-1 Accuracy \\
    \hline \hline
    \multirow{3}{*}{DGC} & \multirow{3}{*}{CIFAR10}  & Baseline     & 93.6\%         \\
    ~          & ~        & Layer wise   & 93.5\%        \\
    ~          & ~        & MergeComp & 93.5\% \\
    \hline
    \multirow{3}{*}{EFSignSGD} & \multirow{3}{*}{ImageNet} & Baseline     & 75.4\%           \\
    ~          & ~        & Layer wise   & 75.2\%              \\
    ~          & ~        & MergeComp & 75.2\%   \\
    \hline
    \end{tabular}
    \vskip -0.1in
    \caption {Top-1 validation accuracy of ResNet50.}
    \label{table:accuray}
    \vskip -0.2in
\end{table}

Table~\ref{table:accuray} shows the Top-1 validation accuracy of ResNet50 on both CIFAR10 and ImageNet with different methods.
Compared with layer-wise compression, \algo can preserve the accuracy of the applied compression algorithms.

\section{Related Work}

There are two directions to improve communication efficiency of large-scale distributed training: 1) higher network bandwidth capacity and 2) smaller communicated data size.
High-bandwidth network devices are equipped to accelerate distributed training. 
For instance, NVLink and NVSwitch~\cite{NVLink} are widely used for intra-machine GPU-to-GPU interconnection and RDMA~\cite{xue2019fast, byteps_osdi} for intra-rack communication.
Unfortunately, even with these advanced hardware, the performance of large-scale distributed training is still far from near-linear scalability because of the large model sizes~\cite{blink, zhang2020network}.
Besides sparsification and quantization algorithms, low-rank compression algorithms~\cite{powersgd, gradzip, idelbayev2020low} are also proposed to reduce the communicated data size.
Local SGD~\cite{local-sgd, woodworth2020local, lin2018don} allows the model to evolve locally on each machine for multiple iterations and then synchronize the gradients. 
It can help compression algorithms further reduce the communication overhead.
Distributed ML frameworks batch multiple tensors for one communication operation to improve the communication performance~\cite{horovod, pytorch_ddp, mxnet, byteps_osdi}, but this mechanism happens after compression and is orthogonal to gradient compression algorithms.
\section{Conclusion}

We propose \algo, a compression scheduler to optimize the scalability of distributed training.
It automatically schedules the compression operations to optimize the performance of compression algorithms without the knowledge of model architectures or system parameters.
Extensive experiments demonstrate that \algo can significantly improve the performance of distributed training over state-of-the-art communication methods without losing accuracy.

\section{Acknowledgement}
We would like to thank the BOLD Lab members for their useful feedback. This research is sponsored by the NSF under CNS-1718980, CNS-1801884, and CNS-1815525.

\bibliography{reference}
\bibliographystyle{icml2020}

\onecolumn
\section{Supplemental Materials}

\begin{assume}
    \label{ass:vg}
   \textup{The variance of gradient among workers is bounded, that is}
    \begin{equation*}
        \small
        \mathbb{E}_{i \sim \mathcal{U}(1, n)} \left\Vert \nabla f_i(x) - \nabla f(x) \right\Vert_2^2 \le \zeta^2 \qquad \forall x,
    \end{equation*}
    \textup{where $\mathcal{U}(1, n)$ is a discrete uniform distribution of integers from 1 to $n$. If all workers share the same training data, $\zeta = 0$.}
\end{assume}

\subsection{Proof to Theorem~\ref{th:sparse}}
\begin{proof}

Refer to a corollary in \cite{jiang2018linear} and we restate it here:

\begin{corollary}
    \cite{jiang2018linear}
    Under Assumptions \ref{ass:lc}-\ref{ass:p} and \ref{ass:vg}, if setting $\gamma=\theta \sqrt{M/K}$ where $\theta > 0$ is a constant, we have the convergence rate for sparsification algorithms as:
    \begin{equation}
        \label{eq:cor1}
        \small
        \frac{1}{K}(\sum\limits_{t=0}^{K-1} \left\Vert \nabla f(\frac{X_t 1_n}{n}) \right\Vert_2^2) \\
        \le \frac{4\theta^{-1}(f(x_0) - f*)+2\theta L\sigma^2}{\sqrt{MK}} + \frac{2pn^2\theta^2L^2\sigma^2 + 6nM\theta^2p^2L^2\zeta^2}{K},
    \end{equation}
    if the number of iterations satisfies $K \ge 12nM\theta^2p^2L^2$.
\end{corollary}

Since we assume all workers share the same training dataset, we can remove the items with $\zeta$ in Inequality~(\ref{eq:cor1}) and have Theorem~\ref{th:sparse}.
    
\end{proof}
\subsection{Proof to Theorem~\ref{th:quantize}}
\begin{proof}

The proof is based on a corollary in \cite{jiang2018linear} and we restate it here:
\begin{corollary}
    \cite{jiang2018linear}
    Under Assumptions \ref{ass:lc}-\ref{ass:p} and \ref{ass:vg}, if using a quantization function with an error bound of $q$ and and setting $\gamma=\theta \sqrt{M/K}$ where $\theta > 0$ is a constant, we have the following convergence rate for quantization algorithms:
    \begin{equation}
    \label{eq:cor2}
        \small
         \frac{1}{K}(\sum\limits_{t=0}^{K-1} \left\Vert \nabla f(x_t) \right\Vert_2^2)
        \le \frac{2\theta^{-1}(f(x_0) - f\kstar)+(1+q)\theta L\sigma^2}{\sqrt{MK}} +  \frac{m}{\sqrt{MK}}\theta qL\zeta^2,
    \end{equation}
    if the number of iterations satisfies $K \ge M\theta^2L^2(1+\frac{q}{n})^2$.
\end{corollary}
Since we assume all workers share the same training dataset, we can remove the items with $\zeta$ in Inequality~(\ref{eq:cor2}) and have the following inequality
 \begin{equation}
    \small
    \begin{split}
    \frac{1}{K}(\sum\limits_{t=0}^{K-1} \left\Vert \nabla f({\rm x}_{i,t}) \right\Vert_2^2)
    & \le \frac{2\theta^{-1}(f({\rm x}_{i, 0}) - f\kstar({\rm x}_i))+(1+q_i)\theta L\sigma^2}{\sqrt{MK}} \\
    & \le \frac{2\theta^{-1}(f({\rm x}_{i, 0}) - f\kstar({\rm x}_i))+(1+q)\theta L\sigma^2}{\sqrt{MK}},
    \end{split}
\end{equation}
where $\rm x_{i, t}$ is $x_i$ at the $t_{th}$ iteration and $q = \max\{q_i\}$.
Since $ \left\Vert \nabla f({\rm x}) \right\Vert_2^2 = \sum\limits_{i=1}^{y} \left\Vert \nabla f({\rm x}_{i}) \right\Vert_2^2$, we have
\begin{equation}
    \label{eq:quant}
    \small
    \begin{split}
    \frac{1}{K}(\sum\limits_{t=0}^{K-1} \left\Vert \nabla f(x_t) \right\Vert_2^2)
    & =  \sum\limits_{i=1}^{y} \frac{1}{K}(\sum\limits_{t=0}^{K-1} \left\Vert \nabla f({\rm x}_{i,t}) \right\Vert_2^2) \\
    & \le \frac{2\theta^{-1}(\sum\limits_{i=1}^{y}f({\rm x}_{i, 0}) - \sum\limits_{i=1}^{y} f\kstar({\rm x}_i))+(1+q)\theta L\sigma^2 y}{\sqrt{MK}} \\
    & \le \frac{2\theta^{-1}(f(x_0) - \sum\limits_{i=1}^{y} f\kstar({\rm x}_i))+(1+q)\theta L\sigma^2 y}{\sqrt{MK}} \\
    \end{split}
\end{equation}

Because $f\kstar$ is the optimal solution, we have $f\kstar \le \sum\limits_{i=1}^{y} f\kstar({\rm x}_i)$ and Inequality~(\ref{eq:quant}) can be rewritten as 
\begin{equation}
    \label{eq:quant}
    \small
    \begin{split}
    \frac{1}{K}(\sum\limits_{t=0}^{K-1} \left\Vert \nabla f(x_t) \right\Vert_2^2)
    \le \frac{2\theta^{-1}(f(x_0) - f\kstar)+(1+q)\theta L\sigma^2 y}{\sqrt{MK}}
    \end{split}
\end{equation}

\end{proof}
\subsection{Proof to Theorem~\ref{th:search}}
\subsubsection{Proof to Lemma~\ref{lemma:space}}
\begin{proof}
    Given a particular $y$, the number of possible partitions with $y$ groups is $\binom{N-1}{y-1}$. Since $y\in [1, N]$, the total number of possible partitions is 
    $\sum_{y=1}^{N} \binom{N-1}{y-1} = 2^{N-1}$.
\end{proof}

\subsubsection{Proof to Lemma~\ref{lemma:y}}
\begin{proof}
Given a partition $X_y = \{x_1, \dots, x_y\}$, the overall compression time is $\sum_{i=1}^{y} h(x_i) = yB_h + \gamma_h D$ and the overall communication time is $\sum_{i=1}^{y} g(x_i) = yB_g + \gamma_g D$, where $D$ is the model size.
These two expressions are both monotonically increasing functions of $y$.
\end{proof}

\subsubsection{Proof to Theorem~\ref{th:search}}
\begin{proof}
    We first prove that $X_2\kstar$ can be solved in $O(\log N)$. 
    Suppose $X_2 = \{x_1, x_2\} = \{x_1, D-x_1\}$, where $D$ is the model size. 
    There are two cases for the overlap time of the first partition group: its communication time is \textit{completely} or \textit{partly} overlapped with the computation of the second group. 
    With a small $x_1$, $g(x_1)$ is completely overlapped and $p(x_1)$ increases with $x_1$.
    At a turning point ($x_1\kstar$), the communication of the first group cannot be finished before the computation of the second group completes so that it comes to the second case. 
    In this case, the communication of the second group can begin right after the communication of the first one and $p(x_1)$ is the computation time of the second group so that $p(x_1)$ decreases with $x_1$.
    According to Lemma~\ref{lemma:y}, both the compression and communication time is constant.
    Therefore, $F(X_2)$ decreases in $(0, x_1\kstar]$ and increases in $(x_1\kstar, D)$.
    Searching for $X_2\kstar$ (i.e., $x_1\kstar$) can be solved with binary search in $O(\log N)$.
    
    When $y>2$ and suppose $X_y = \{x_1, \dots, x_y\}$, we can first fix $x_1, \dots, x_{y-2}$ and then solve the optimal $x_{y-1}$ to minimize $F(X_y)$.
    There are three cases for the communication of $x_{y-2}$: 1) $g(x_{y-2})$ is not overlapped; 2) $g(x_{y-2})$ is partly overlapped and 3) $g(x_{y-2})$ is completely overlapped.
    For cases 1) and 2), the value of $x_{y-1}$ will not affect $F(X_y)$ because there is no overlap time for $x_{y-1}$ and $x_y$; for case 3), the optimal $x_{y-1}$ can be solved in $O(\log N)$ based on the above analysis.
    Since $\binom{N-2}{y-2} \log N < N^{y-2}\log N$, $X_y\kstar$ can be solved in $O(N^{y-2}\log N)$.
    Therefore, Algorithm~\ref{alg:partition} can solve all optimal partition $X_i\kstar$, where $2 \le i \le Y$, in $O(N^{y-2}\log N)$ and return the best $X_i\kstar$.
\end{proof}


\end{document}